\documentclass{jfm}
\usepackage{graphicx}
\usepackage{subcaption}
\usepackage{epstopdf}
\usepackage{natbib}
\usepackage[pdftex,colorlinks]{hyperref}
\usepackage{amsmath,amsfonts,amssymb,amsbsy,amscd,amsgen}
\usepackage{mathrsfs}

\newcommand\bs{\ensuremath\boldsymbol}
\newcommand{\eqnref}[1]{(\ref{#1})}

\usepackage{booktabs}
\usepackage{multicol}
\usepackage{multirow}
\usepackage{tabularx}
\usepackage{wasysym}
\usepackage{enumerate}
\usepackage{enumitem}
\setenumerate{itemsep=3pt,topsep=3pt}

\newcommand{\rank}[2]{$\Pi_{#1} (\mathrm{EQ}#2)$}

\shorttitle{New equilibrium solution branches discovered using a project-then-search method}
\shortauthor{M. A. Ahmed and A. S. Sharma}

\title{New equilibrium solution branches of plane Couette flow discovered using a project-then-search method}
\author{M. A. Ahmed\aff{1},
	\and A. S. Sharma\aff{1,}\aff{2}
	\corresp{\email{a.sharma@soton.ac.uk}}}

\affiliation{\aff{1}Department of Aerospace Engineering, University of Southampton, Southampton, SO17 1BJ, UK
\aff{2}Kavli Institute for Theoretical Physics, University of California, Santa Barbara, CA 93106, USA}

\graphicspath{{./figs/}}

\begin{document}
\maketitle
\begin{abstract}
	New equilibrium solution branches for plane Couette flow are reported which add to the inventory of known solution branches.
    The exact solutions are found by projecting known equilibria onto the resolvent modes of McKeon \& Sharma (\emph{J.~Fl.~Mech.}, vol.~658, 2010, pp.~336-382) to generate approximate solutions that are subsequently used as seeds in a Newton-Krylov-hookstep search.
    Searches initialised with these projections have a convergence rate of $96\%$, 
    leading to the discovery of 16 new equilibria.
	The low-dimensional nature of the resulting equilibria is attributed to the projection generated by the resolvent model; 
	resolvent modes for a given equilibrium solution span a low-dimensional space which the solution approximately inhabits.
	This property is exploited to generate new branches of equilibria in plane Couette flow, and to jump to known branches.
	The new branches include bifurcations from previously known bifurcation curves and a disconnected bifurcation curve which displays interesting behaviour when continued in Reynolds number.
	The unstable manifolds of the new exact coherent states are also computed to expand the known state space structure of plane Couette flow.
\end{abstract}

\section{Introduction}
A complete understanding of turbulent flows is one of the most important and fascinating problems in classical physics.
The problem has been looked at from various perspectives; here we adopt the nonlinear dynamical systems standpoint.
Wall-bounded turbulent flows contain persistent macroscopic structure that recurs in space and time known as \emph{coherent structures}. 
\cite{Kline1967} performed experiments which showed spatially organised streaks in the turbulent boundary layer.
Further work has revealed that there is order within apparently random turbulence \citep{Adrian2007,Cantwell1981,Holmes1998,Robinson1991,Sharma2013}.
\cite{Hamilton1995} identified a well-defined quasi-cyclic process in plane Couette flow at low Reynolds number.
This quasi-cyclic process eventually manifested in the `Self-Sustaining Process' theory (SSP) \citep{Waleffe1997} that describes the roll-streak behaviour in terms of forced response of streaks to rolls, growth of streak instabilities, and nonlinear feedback from streak instabilities to rolls.
It is widely thought that similar processes are important in turbulent flow.

The dynamical systems picture can be traced back to \cite{Hopf1948}, who proposed that turbulence is described by a finite-dimensional manifold invariant under the action of the state space flow.
In this picture, equilibria are fixed points that give structure to the invariant manifold, and their bifurcation structure determines behaviour across Reynolds number. In a sense, all work on turbulence is ultimately aimed at producing a concise description of this invariant manifold.

The idea is that turbulent flow can be described through invariant solutions of the NSE and their connecting orbits.
There are four classes of such solution that we are concerned with.
Equilibria, which are constant in time; relative equilibria (also called travelling wave solutions), which are constant in a co-moving frame of reference;
periodic orbits, which repeat over a fixed time period; and relative periodic orbits, which are periodic orbits in a co-moving frame of reference.

For these reasons, \cite{Kawahara2012} pose the finding of sufficiently many invariant solutions to fully describe a turbulent flow as one of the outstanding problems in the field.
Unfortunately, with current methods, computing invariant solutions involves trial and error and computational searches often fail.
In what follows, we focus on the simplest objects, equilibria, and show how
many new equilibria may be generated cheaply from those already known.

% Methods for generating initial conditions
Trivial homogeneous equilibrium solutions of the NSE, such as the stable laminar flow state in plane Couette and Poiseuille flow, can be found with very little effort; these solutions are easy to find analytically as well as numerically. 
Less symmetric solutions on the other hand are more difficult to find.
The most commonly used method for finding invariant solutions is the Newton search because it converges quadratically.
However, as typically implemented the convergence is only local; to guarantee convergence an initial guess that is close to a solution must be provided.
Finding suitable initial states for the algorithm is the tricky task.
During the course of this investigation \cite{Farazmand2016} developed a hybrid adjoint-Newton algorithm that provides global convergence from any given initial condition to find equilibria of two-dimensional Kolmogorov flow.

There are various methods that can be used to generate initial guesses for searches, the most widely used of which we list below.
\begin{enumerate}[wide = 0pt,align = right, labelwidth = 2em, leftmargin =\dimexpr\labelwidth + \labelsep+1em, font=\normalfont, label=(\alph*), itemsep=1em, labelsep=1em]
	\item Bisection:\\
	This approach involves varying the initial amplitude of a velocity field depending on its temporal evolution.
	The time-evolved flow field which sits on the boundary between the laminar and turbulent states (a flow field that neither laminarises nor becomes fully turbulent, a so-called edge state) is used as an initial guess for a Newton search.
    This method is used by \cite{Duguet2008b, Duguet2009, Hwang2016} and \cite{Avila2013}.
    A related but more sophisticated method, which modifies the Reynolds number with on-line feedback control to automatically achieve the same effect, is used by \cite{Willis2017}.
	
	\item Homotopy:\\
	In this method a convergence control parameter is introduced and solutions are found as the parameter changes via continuation. 
	\cite{Waleffe2003} describes it as smoothly deforming the base flow into the desired flow while tracking the solutions with Newton\textsc{\char13}s method.
	Several studies have successfully used this approach to build solution connections between Taylor-Couette, plane Couette and plane Poiseuille flow, 
	see \cite{Nagata1990, Nagata1997, Nagata1998}, \cite{Faisst2003} and \cite{Waleffe2001, Waleffe2003}.
	
	\item Recurrence plots:\\
	This technique involves studying the time-evolution of a perturbed flow field for periodic patterns.
	The most dynamically important fields are then used as initial guesses in a Newton search.
	The method has been successfully used to find equilibria, travelling waves and periodic orbits of the NSE, see \cite{Chandler2013, Cvitanovic2010, Gibson2009}.
	For a comprehensive review of the technique refer to the work of \cite{Marwan2007}.

	\item Projections:\\
    Introduced here, low-rank projections of known equilibrium solutions are used as initial guesses in the search for other solutions.
    The projected velocity field resembles its parent and maintains its dominant flow characteristics (in the sense of the norm used to define the projection).
	The projected fields have structures derived from known solutions, making them good initial guesses for the search algorithm.
\end{enumerate}

% Nagata was the first
The first to discover non-trivial high-dimensional nonlinear solutions of the NSE at moderate-Reynolds number was \cite{Nagata1990}, who computed a pair of unstable three-dimensional equilibria in plane Couette flow using homotopy and bifurcations from a wavy vortex solution of Taylor-Couette flow.
Continuing the two solutions downwards in Reynolds number reveals that they originate from a saddle-node bifurcation and contain a wavy low-velocity streak flanked by counter-rotating vortices; 
the upper branch solutions consist of weak streaks with strong vortices whereas the lower branch solutions consist of weak vortices but stronger streaks.
Their structure persists at higher Reynolds numbers and resembles coherent structures observed in the near-wall region of wall-bounded turbulent flows \citep{Jeong1997, Stretch1990}.
The same solutions were found independently by \cite{Clever1997} and \cite{Waleffe1998, Waleffe2003}.
\cite{Itano2009}, \cite{Nagata1997}, \cite{Gibson2008,Gibson2009} and \cite{Schmiegel1999} found a multitude of equilibria in plane Couette flow that are not related to \cite{Nagata1990}'s solutions but exhibit similar flow structures.
% Other solutions in other geometries.
Relative equilibria have been found using the homotopy approach in plane Poiseuille flow \citep{Ehrenstein1991,Itano2001},
pipe flow \citep{Faisst2003,Wedin2004} and 
square ducts \citep{Okino2010,Uhlmann2010,Wedin2009}.
For a comprehensive review refer to \cite{Kawahara2012}.

% Visualising EQ in state space
\cite{Gibson2008} developed a representation-independent method of visualising invariant solutions in state space, the details of which are given in \S\ref{sec:Statepace}.
\begin{figure}
	\centering
	\includegraphics[width=0.7\textwidth]{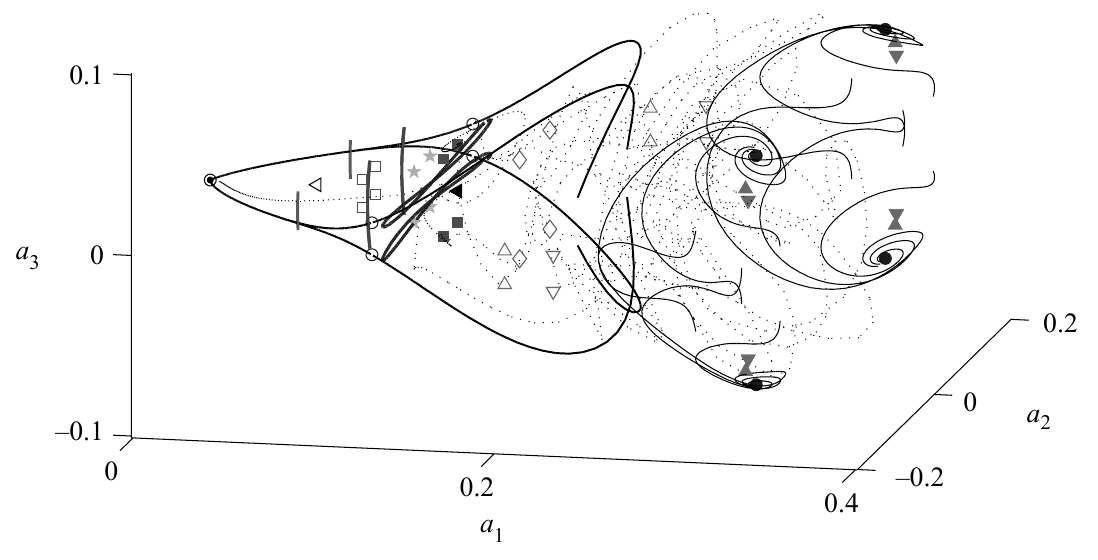}
	\caption{A three-dimensional state space portrait of plane Couette flow showing previously known equilibria (symbols), their manifolds (solid lines) and a turbulent trajectory (dashed line).
	Figure taken from \cite{Gibson2009}; see original source for a detailed description.}
	\label{fig:gibson_state_space}
\end{figure}
Figure \ref{fig:gibson_state_space} depicts the state space portrait that \citeauthor{Gibson2008} constructed.
The portrait allows us to follow a transitional or turbulent trajectory as it wanders near fixed points and is directed by the unstable manifolds of the equilibria.
This state space representation organises solutions relative to the laminar solution thus revealing the structure of the laminar attractor. 
Additionally, the portrait helps in seeing heteroclinic connections between fixed points as shown by \cite{Gibson2008}, who were the first to find a heteroclinic connection between two non-trivial equilibria.
\cite{Halcrow2008} found further heteroclinic connections between equilibria and demonstrated that these connections are pertinent to obtaining a global picture of turbulent flow dynamics in state space.

% Link between high- & low-dimensionality
It is claimed that the connecting orbits between invariant solutions of the NSE can be used to describe low-dimensional processes embedded within fully developed turbulent flows \citep{Halcrow2008a}.
The high-dimensional dynamics of turbulence contains macroscopic organised structure that looks like and shares mechanisms with invariant solutions.
For this reason, \cite{Waleffe2001} referred to the exact solutions of the NSE as \emph{exact coherent structures}.
The solutions closely resemble coherent structures seen in experiments and direct numerical simulations (DNS), but they lack the complex spatio-temporal intermittency characteristics of coherent structures observed in turbulent flows.
Therefore, understanding exact coherent structures and their connections can clarify the fundamental dynamics that form the backbone of fully turbulent flows.
\cite{Jimenez2001} showed that the underlying dynamics in the near-wall region reveal themselves in intermittent and deterministic processes that are low-dimensional and \cite{Holmes1998} argue that low-dimensional processes are embedded in global high-dimensional fully developed turbulence.
From the nonlinear dynamical systems viewpoint, connecting orbits can describe these low-dimensional processes.
With this in mind, we provide structure to the laminar attractor first.

The structure of the paper is as follows.
The project-then-search methodology is described in \S\ref{sec:Methodology}.
The classification of symmetries in plane Couette flow that support equilibria are given in \S\ref{sec:Symms}.
The new equilibria are presented and discussed in \S\ref{sec:Results} and \S\ref{sec:Discussion}.
Conclusions are given in \S\ref{sec:Conclusions}.

\section{Methodology}\label{sec:Methodology}
\subsection{Plane Couette flow}\label{sec:PCF}
\begin{figure}
	\centering
	\includegraphics[width=.5\linewidth]{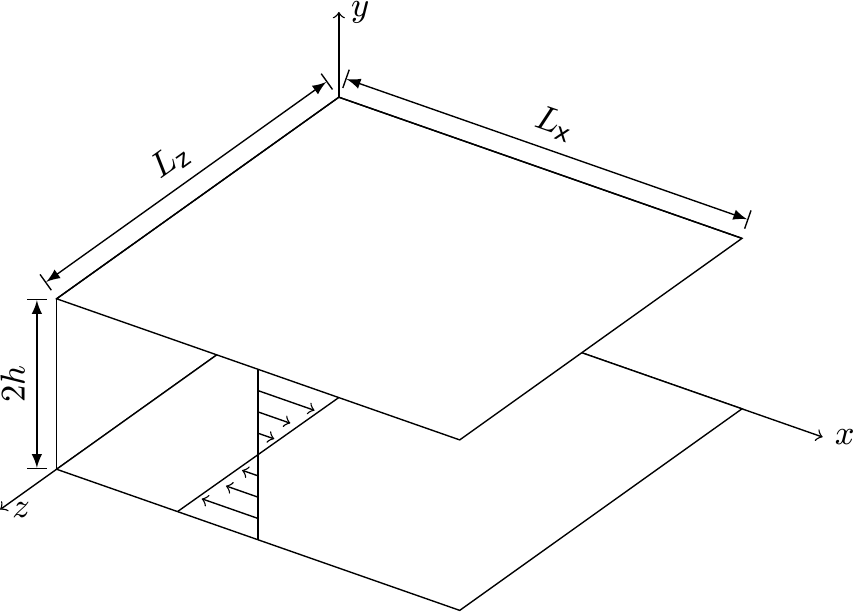}
	\caption{Channel geometry for plane Couette flow. 
		Periodic boundary conditions are assumed in the streamwise ($x$) and spanwise ($z$) directions. 
		The channel half-height is restricted to $h=1$, such that $y \in [-1,1]$.
		The boundaries move in-plane in opposite directions at a speed of $\pm1$ in the streamwise direction.
		A no-slip boundary condition is imposed at the walls, such that $\bs{u}(x,\pm1,z)=[\pm1,0,0]$.}
	\label{fig:01}
\end{figure}

The flow geometry (see figure \ref{fig:01}) and all parameters are chosen to be consistent with the work of \cite{Gibson2009}.
% Equations of motion
The nondimensional NSE for an incompressible fluid are
\begin{subeqnarray}\label{eqn:NSE}
	\frac{\partial \bs{u}}{\partial t} + \bs{u} \cdot \nabla \bs{u} & = & - \nabla  p  + Re^{-1}\nabla^{2}\bs{u},\\
	\nabla \cdot \bs{u} & = & 0,
\end{subeqnarray}
where $\bs{u}(x,y,z,t) = [u\ v \ w]^{T}$ is the velocity vector in the streamwise $x$, wall-normal $y$ and spanwise $z$ directions,\, $t$ is time,\, $ p (x,y,z,t)$ is the pressure, $\nabla$ is the gradient operator, $\nabla^{2}$ is the Laplace operator and $Re$ is the Reynolds number. 
The Reynolds number is defined as $Re = Uh/\nu$,  where $U$ is half the relative velocity of the plates, $h$ is the channel half-height and $\nu$ is the kinematic viscosity.

% Domain.
%The domain is defined as $\Omega = [0,L_{x}]\times[-1,1]\times[0,L_{z}]$.
The domain has periodic boundary conditions in the streamwise and spanwise directions.
Spatial periodicity is specified in terms of fundamental streamwise and spanwise Fourier wavenumbers, $\alpha$ and $\beta$, respectively.
The relation between the spatial wavenumbers and the domain is given as $L_{x} = 2\pi m/\alpha$ and $L_{z} = 2\pi n/\beta$ where $m, n \in \mathbb{Z}^{+}$.
The equilibria are computed on the domain $\Omega = [\frac{2\pi}{1.14}, 2, \frac{2\pi}{2.5}]$ which is discretised onto a $[32, 35, 32]$ grid.
In this study the pressure gradient is fixed at zero and integration forward in time is performed with time step $\Delta t = 0.03125$.

% Velocity breakdown and E,D.
The unit vectors in the $x,y,z$ directions are denoted $\hat{\bs{x}},\hat{\bs{y}},\hat{\bs{z}}$ and  the plane Couette base flow is defined as $y\hat{\bs{x}}$.
Hence, the total velocity is defined as $\bs{u} = \tilde{\bs{u}} + y\hat{\bs{x}}$, where $\tilde{\bs{u}}$ is the velocity difference from laminar.
In the present study the $\mathscr{L}^{2}$-inner product and norm are defined as 
\begin{subequations}\label{eqn:IP_norm}
	\begin{align}
		\langle \bs{a}, \bs{b} \rangle & = \frac{1}{2L_{x}L_{z}}\int_{\Omega}\; \bs{a}\cdot\bs{b} \;dx\ dy\ dz \\
		\| \bs{a}\|^{2} & = \langle \bs{a}, \bs{a} \rangle.
	\end{align}
\end{subequations}
The energy is defined as $\| \tilde{\bs{u}}\| = \sqrt{\langle \tilde{\bs{u}}, \tilde{\bs{u}} \rangle}$, the kinetic energy density as $E = \frac{1}{2} \| \bs{u}\|^{2}$ and the dissipation rate as $D = \| \nabla \times \bs{u}\|^{2}$.
The role that certain solutions play in turbulent dynamics can be interpreted from the dissipation rate, e.g. a lower dissipation rate means that the equilibrium is far from the turbulent attractor and closer to laminar flow.

\subsection{Computational method for search}\label{sec:Comp_Method}
The new equilibria found in this report are derived from equilibria previously discovered by \cite{Nagata1990}, \cite{Gibson2008,Gibson2009} and \cite{Halcrow2008}, which can be found at \url{channelflow.org}.
The open source library \texttt{Channelflow} was used to find, continue and analyse all solutions in our investigation \citep{channelflow}.
All computational settings are the same as in the work of \cite{Gibson2009}.

%% Search algorithm summary (NKH)
In the present work, the Newton-Krylov-hookstep (NKH) search algorithm developed by \cite{Viswanath2007} is used to find invariant solutions of the NSE, for a detailed explanation of the algorithm refer to the work of \cite{Halcrow2008a} and \cite{Viswanath2007}.
The algorithm finds approximate solutions to the following equation,
\begin{equation}\label{eqn:NKH}
	G(\tilde{\bs{u}}, \sigma, t) = \sigma f^{t}(\tilde{\bs{u}}) - \tilde{\bs{u}} = 0,
\end{equation}
where $f^{t}(\tilde{\bs{u}})$ is a time-mapped instance of the initial flow field at time $t$, $\tilde{\bs{u}}$ is the initial flow field and $\sigma$ is the symmetry of the flow field (see \S\ref{sec:Symms}).
The convergence criteria for the searches is $ \| G\|  \leq 10^{-13}$.
An equilibrium solution is defined as $\tilde{\bs{u}}(x,y,z,t) = \tilde{\bs{u}}_{EQ}(x,y,z)$, 
therefore when there is close to no difference between the time-evolved state and the initial flow field an equilibrium solution is deemed to have been found.
For the searches performed in the current work, the NKH algorithm is unconstrained with respect to symmetries, 
only the final solution is inspected for symmetries.

% Accuracy computation
Following \cite{Gibson2009} and \cite{Viswanath2007}, to determine the accuracy of a solution it is interpolated from a grid resolution of $[32, 35, 32]$ onto a $[48, 49, 48]$ grid and then time integrated for $T=1$ with $dt=0.02$.
Then the accuracy is determined by calculating the residual, defined as
\begin{equation}\label{eqn:acc}
	\frac{\| f^{T=1}(\tilde{\bs{u}}) - \tilde{\bs{u}} \|}{\| \tilde{\bs{u}} \|}.
\end{equation}
All searches for new equilibria and their bifurcation curves were also performed at a higher grid resolution of $[48,65,48]$ to ensure that the searches were well-resolved; 
all results were found to agree with those found at the lower grid resolution of $[32, 35, 32]$ down to the fifth significant figure.
Invariant solutions have been found in the context of minimal flow units 
(small periodic domains just large enough to sustain turbulent flow or contain a single coherent structure, \cite{Jimenez1991}).

\subsection{Low-rank projections}\label{sec:LRP}
The practical difficulty with using Newton's method to find equilibria is that the search may fail or return the trivial laminar solution. 
The aim here is to generate initial guesses that are close enough to new equilibria to converge. 
As such, we generate such initial guesses by projecting known equilibria onto resolvent modes.
The resolvent model is used to provide a physically relevant, ordered basis in which the velocity field can be expanded. 
In principle, projections onto other bases could be used.

To this end, a short summary of the pertinent aspects of the resolvent model and associated projection is provided below. 
For a detailed derivation of the model in the context of turbulence, the reader may refer to \cite{McKeon2010}, and for the projection step, to \cite{Sharma2016}.

It was previously shown in \citet{Sharma2016} that these projections are often close to the original solutions, even when the projection is very low rank. 
We will see that the projections are also often close to other, nearby, solutions.

Consider a long-time solution to \eqref{eqn:NSE}, $\bs{u}$. This solution may be expanded into its harmonic and transient parts,
\begin{equation}
	\bs{u}(t) = \frac{1}{2\pi} \int_{-\infty}^{\infty} e^{i \omega t} \hat{\bs{u}}(\omega) d \omega + \bs{T}(t)
\end{equation}
where the dependence on $x, y, z$ has been suppressed.
For the case where flow has already decayed onto the attractor, or for any recurrent flow, $\bs{T}=0$. 
Notice that the temporal mean is associated with $\hat{\bs{u}}(0)$.

Writing \eqref{eqn:NSE} as $\partial \bs{u}/\partial t = \bs{f}(\bs{u})$, 
an expansion about the temporal mean $\bar{\bs{u}} := \hat{\bs{u}}(\omega = 0)$ gives
\begin{subequations}
	\begin{align}
		\bs{u} &= \bar{\bs{u}} + \tilde{\bs{u}} \\
		\frac{\partial \tilde{\bs{u}}}{\partial t} &= \left. \frac{\partial \bs{f}}{\partial \bs{u}} \right\vert_{\bar{\bs{u}}}  \tilde{\bs{u}}(t) 
		+ \mathcal{O}(2) \\
		&= \mathcal{L} \tilde{\bs{u}}(t) + \tilde{\bs{g}}(t)
	\end{align}
\end{subequations}
where $\tilde{\bs{g}}$ represents the second-order terms in the expansion of $\bs{f}$ (the Reynolds stress gradients).
Note that we are performing an expansion around the mean of a specific equilibrium solution, rather than the turbulent mean, as in the earlier work.
Similarly expanding $\bs{g}$ in its Fourier coefficients and rearranging gives
\begin{equation}\label{eqn:resolvent}
	\hat{\bs{u}}(\omega) = \left( \omega I - \mathcal{L} \right)^{-1} \hat{\bs{g}}(\omega),
\end{equation}
at any $\omega \neq 0 $.
In this formulation, the second-order terms act to excite the state rather than being truncated.

The operator $\mathcal{H}(\omega) = \left( \omega I - \mathcal{L} \right)^{-1}$ is the resolvent of $\mathcal{L}$ and is a linear mapping from the Reynolds stress gradients to the velocity field.
The idea of the projection step is to find the optimal projection $\Pi(\omega)$ of rank $M$ that approximates $\mathcal{H}(\omega)$.
Since the resolvent operator is linear, the optimal projection is provided by the singular value decomposition (SVD) of $\mathcal{H}$.

Noting that the SVD of $\mathcal{H}$ induces a Fourier decomposition in the spatially invariant directions ($x$ and $z$, see \cite{Sharma2016b}), 
it is profitable to perform the SVD separately at each Fourier frequency-wavenumber combination $K=(\alpha,\ \beta)$ for the Fourier representation of \eqref{eqn:resolvent},
\begin{subequations}
	\begin{align}
		\mathcal{H}_K(\omega)\bs{a} &= \left( \omega I - \mathcal{L}_K \right)^{-1}\bs{a}\\
		&= \sum_{m=1}^\infty \bs{\psi}_K^m(\omega) \sigma_K^m(\omega) \left< \bs{\phi}_K^m(\omega), \bs{a}\right>_y.
	\end{align}
\end{subequations}
The SVD has the useful orthogonality properties
\begin{subequations}
	\begin{align}
		\left<\bs{\phi}_K^m(\omega), \bs{\phi}_K^{m'}(\omega)\right> &= \delta_{m,m'}\\
		\left<\bs{\psi}_K^m(\omega), \bs{\psi}_K^{m'}(\omega)\right> &= \delta_{m,m'}\\
		\sigma_1 \geq \sigma_2 \geq \ldots &\geq \sigma_m \geq \ldots
	\end{align}
\end{subequations}
In this case, the $\bar{\bs{u}}$ to be used in forming $\mathcal{L}$ is the $\alpha=0,\ \beta=0$ component of the temporal mean.

The velocity field may then be expressed as an expansion in resolvent modes,
\begin{equation}\label{eqn:resolvent_expansion}
	\bs{u}(x,y,z,t) = \frac{1}{2\pi} \sum_{\alpha,\ \beta} \int_{-\infty}^{\infty} e^{i (\omega t + \alpha x + \beta z) } \sum_{m=1}^\infty \bs{\psi}_K^m (\omega, y) c_K^m d \omega .
\end{equation}

Applying the optimal rank-$M$ projection $\Pi_M$ gives
\begin{align}
	\bs{u} = \Pi_M \bs{u} + \bs{u}^\perp 
\end{align}
with the sum over $m$ in \eqref{eqn:resolvent_expansion} being split as
\begin{align}
	\Pi_M \bs{u} &= \frac{1}{2\pi} \sum_{\alpha,\ \beta} \int_{-\infty}^{\infty} e^{i (\omega t + \alpha x + \beta z) } \sum_{m=1}^M \bs{\psi}_K^m (\omega, y) c_K^m d \omega , \\
	\bs{u}^\perp &= \frac{1}{2\pi} \sum_{\alpha,\ \beta} \int_{-\infty}^{\infty} e^{i (\omega t + \alpha x + \beta z) } \sum_{m=M+1}^\infty \bs{\psi}_K^m (\omega, y) c_K^m d \omega . \\
\end{align}

The idea of the projection is that the flow resides mostly (in an energy sense) in the subspace that $\Pi_M$ projects onto.
For an explanation in terms of Koopman modes, refer to \citet{Sharma2016b}.
The projection $\Pi_M$ is calculated for each known solution and then used as an initial guess for the NKH search.

\section{Symmetries in plane Couette flow}\label{sec:Symms}
A symmetry operation $\sigma$ is a linear transformation of the state of a dynamical system which commutes with integration forward in time,
\begin{equation}
	\sigma \dot{\tilde{\bs{u}}} = \sigma f(\tilde{\bs{u}}) = f(\sigma \tilde{\bs{u}}).
\end{equation}
We define an isotropy group of $\tilde{\bs{u}}$ as a group that contains all symmetries that satisfy $\sigma \tilde{\bs{u}} = \tilde{\bs{u}}$.

Solutions in an equivariant system (such as plane Couette flow, which is highly symmetric) can satisfy all of the system's symmetries, a subgroup of the symmetries or none of the symmetries.
Typically, a turbulent trajectory has no symmetries, i.e. its isotropy group consists of the identity operation $\{e\}$ only.
The laminar solution in plane Couette flow obeys every continuous symmetry that the geometry allows (see figure \ref{fig:01}), this symmetry group is defined as $\Gamma$ (see \cite{Halcrow2008a} for full derivation).

% What are they for PCF?
For plane Couette flow, the NSE are invariant under reflection in the $yx$-plane, rotation about the $z$-axis by $\pi$, pointwise-inversion through the origin and continuous translations in the $x$ and $z$ axes.
Accordingly isotropy groups of the equilibria reported here contain combinations of reflection, rotation, pointwise-inversion and translation, defined as
\begin{subequations}\label{eqn:Continuous_Symms}
	\begin{align}
		\sigma_{z} & : [u, v, w](x, y, z) \rightarrow [u, v, -w](x, y, -z)\\
		\sigma_{x} & : [u, v, w](x, y, z) \rightarrow [-u, -v, w](-x, -y, z)\\
		\sigma_{xz} & : [u, v, w](x, y, z) \rightarrow [-u, -v, -w](-x, -y, -z)\\
		\tau(\delta x, \delta z) & : [u, v, w](x, y, z) \rightarrow [u, v, w](x + \delta x, y, z + \delta z),
	\end{align}
\end{subequations}
respectively.

% The TW conundrum
The symmetries that a particular solution adheres to dictate the type of solution it is.
The reflection symmetry reverses the spanwise velocity, $w$, therefore any solution that is invariant under $\sigma_z$ cannot be a spanwise travelling wave.
Similarly, the rotation symmetry reverses the streamwise velocity, $u$, therefore for any solution that is invariant under $\sigma_x$ does not permit a streamwise travelling wave.
Consequently, if a solution is invariant under $\sigma_{xz}$ it cannot be a travelling wave solution in $x$ or $z$.
This implies that the solutions that obey $\sigma_{xz}$ are equilibria since they are spatially static.
All of the equilibria in this study satisfy $\sigma_{xz}$, since the equilibria are derived from the solutions found by \cite{Nagata1990} and \cite{Gibson2009}, who sought equilibria that obeyed $\sigma_{xz}$.

% The discrete shifts
The periodic boundary conditions impose discrete translation symmetries on the equilibria.
If a field is fixed under a discrete shift $\tau(L_x/n,0)$, it is periodic on the smaller spatial domain $x \in [0, L_x/n], n \in \mathbb{Z}^{+}$, similarly for $z$.
Using half-cell shifts in the streamwise ($\varDelta x = L_x/2$) and spanwise ($\varDelta z = L_z/2$) directions only, the following symmetry operations can be defined
\begin{subequations}\label{eqn:symms}
	\begin{align}
		\sigma_{x} =\; & \theta_1 [u, \; v, \; w](x, \;y, \;z) = [-u, \; -v, \; w] (-x, \;-y, \;z),\\
		\tau_z\sigma_{z} =\; & \theta_2 [u, \; v, \; w](x, \;y, \;z) = [u, \; v, \; -w] (x, \;y, \;-z + \varDelta z),\\
		\tau_z\sigma_{xz} =\; & \theta_3 [u, \; v, \; w](x, \;y, \;z) = [-u, \; -v, \; -w] (-x, \;-y, \;-z + \varDelta z),\\
		\tau_x\sigma_{z} =\; & \theta_4 [u, \; v, \; w](x, \;y, \;z) = [u, \; v, \; -w] (x + \varDelta x, \;y, \;-z),\\
		\tau_x\sigma_{x} =\; & \theta_5 [u, \; v, \; w](x, \;y, \;z) = [-u, \; -v, \; -w] (-x + \varDelta x, \;-y, \;-z),\\
		\tau_{xz} =\; & \theta_6 [u, \; v, \; w](x, \;y, \;z) = [u, \; v, \; w] (x + \varDelta x, \;y, \;z + \varDelta z),\\
		\tau_{xz}\sigma_{x} =\; & \theta_7 [u, \; v, \; w](x, \;y, \;z) = [-u, \; -v, \; w] (-x + \varDelta x, \;-y, \;z + \varDelta z),
	\end{align}
\end{subequations}
These operations are used to define the isotropy groups to which all solutions in our work belong. 
For more details on the other isotropy groups of plane Couette flow see \cite{Halcrow2008a}.
% Present symmetry operations available
All equilibrium solutions found in the present study belong to the following one of the following isotropy subgroups,
\begin{subequations}\label{eqn:isotropy_groups}
	\begin{align}
		\Theta & = \{e, \;\theta_1, \;\theta_2, \;\theta_3, \;\theta_4, \;\theta_5, \;\theta_6, \;\theta_7\},\\
		K & = \{e, \;\theta_1, \;\theta_2, \;\theta_3\},\\
		\Sigma & = \{e, \;\theta_4, \;\theta_7, \;\theta_3\},\\
		\Theta_3 & = \{e, \;\theta_3\}.	
	\end{align}
\end{subequations}
Here $\Theta_3 \subset \Sigma, K \subset \Theta \subset \Gamma$. 
The isotropy subgroup $K$ only admits translations in the spanwise direction; an equilibrium solution that belongs to this symmetry subgroup was discovered and is detailed in Section \ref{sec:eq_props}.
The isotropy subgroup $\Sigma$ is called $S$ in the works of \cite{Nagata1990}, \cite{Gibson2009} and \cite{Waleffe2003}.
$\Theta$ can also be expressed as $\Theta = \Sigma \times \{e,\tau_{xz}\}$. 
These are only some of the symmetry subgroups of $\Gamma$. 
Other subgroups might also play an important role in turbulent dynamics;
there may be other equilibria that obey different symmetries, or none at all.

% SPACES
The full space of velocity fields for plane Couette flow can be written as
\begin{equation}
	\begin{aligned}
		\mathbb{U} = \{\bs{u} \in L^2(\Omega)\;|\; &\nabla \cdot \bs{u} = 0,\; \\ 
		&\bs{u}(x,\pm1,z) = 0,\; \\ 
		&\bs{u}(x,y,z)=\bs{u}(x+L_x,y,z)=\bs{u}(x,y,z+L_z) \}.
	\end{aligned}
\end{equation}
The isotropy subgroups can be used to define invariant subspaces of velocity fields that are invariant under symmetry operations contained within the isotropy subgroups.
For an arbitrary isotropy subgroup $H$, the $H$-invariant subspace can be defined as 
\begin{equation}
	\mathbb{U}_{H} = \{\ \bs{u} \in \mathbb{U} \ |\ h\bs{u}=\bs{u},\ \forall\; h \in H\ \},
\end{equation}
where $h$ is a symmetry operation.
$H$ can be replaced with any of the isotropy subgroups defined in (\ref{eqn:isotropy_groups}).

\subsection{Geometry of plane Couette state space}\label{sec:Statepace}
Plane Couette flow dynamics can be represented as state space portraits.
\cite{Gibson2008} developed a basis that defines a representation independent state space which shows the relationships between equilibria and allows us to chart turbulent trajectiories.
A brief summary of the state space visualisation method developed by \citeauthor{Gibson2008} is given below.

The premise of the method is that velocity fields may be projected onto an orthonormal basis,
\begin{equation}\label{eqn:basis_projection}
	a_{n}(t) = \langle \bs{u}(t), \bs{e}_n \rangle,
\end{equation}
where $\bs{e}_n$ is a unit basis vector and $a_n$ is the low-dimensional projection of a given velocity field $\bs{u}(t)$.
Therefore, a state space trajectory is projected onto the $\{\bs{e}_n\}$ coordinate frame.

% How/Which orthonormal basis is chosen?
Following the work of \cite{Gibson2008}, an orthonormal translational basis is constructed based on streamwise and spanwise half-domain shifts of \cite{Nagata1990}'s upper branch solution ($\bs{u}_{\mathrm{EQ2}}$), defined as
\begin{equation}\label{eqn:state_space}
	\begin{matrix}
		 & & \tau_{x} & \tau_{z} & \tau_{xz}\\
		\bs{e}_1 = \gamma_1 (1 + \tau_x + \tau_z + \tau_{xz})\bs{u}_{\mathrm{EQ2}} & & S & S & S \;,\\ 
		\bs{e}_2 = \gamma_2 (1 + \tau_x - \tau_z - \tau_{xz})\bs{u}_{\mathrm{EQ2}} & & S & A & A \;,\\ 
		\bs{e}_3 = \gamma_3 (1 - \tau_x + \tau_z - \tau_{xz})\bs{u}_{\mathrm{EQ2}} & & A & S & A \;,\\ 
		\bs{e}_4 = \gamma_4 (1 - \tau_x - \tau_z + \tau_{xz})\bs{u}_{\mathrm{EQ2}} & & A & A & S \;,
	\end{matrix}
\end{equation}
where $\gamma_n$ is a normalisation constant such that $\|\bs{e}_n\|=1$.
Here, $\tau_i$ represents a half-domain shift in the direction specified by the subscript $i$.
% Symmetries of basis functions
On the right hand side the last three columns denote the symmetry of each basis function under the appropriate translation, e.g $A$ in the $\tau_{x}$ column means that $\tau_{x}\bs{e}_{n} = -\bs{e}_{n}$ (anti-symmetric) and $S$ means $\tau_{x}\bs{e}_{n} = \bs{e}_{n}$ (symmetric).
% What is the origin then?
The origin in this state space is the laminar solution $\bs{u}_{\mathrm{EQ0}}$ since it is invariant under all symmetries and all solutions presented here are expressed as are differences from laminar.
As emphasised by \cite{Gibson2008}, this orthonormal basis definition is one of many.

It should be noted that a basis can be constructed from any field as there is no pre-determined method of selecting a fluid state, being at the author's discretion.
Following \cite{Gibson2008}, we chose to select an orthonormal basis formed from $\bs{u}_{\mathrm{EQ2}}$.

\section{Results}\label{sec:Results}
% Section description
Firstly, the energy and symmetry properties of all equilibria collated in table \ref{tab:equilibria} are described and compared in \S\ref{sec:eq_props}.
Secondly, we visualise the new equilibria and their invariant manifolds in state space (defined in \eqnref{eqn:state_space}) in \S\ref{sec:eq_state_space}.
Thirdly, the equilibria are continued in $Re$ and the resulting bifurcation curves are analysed in \S\ref{sec:eq_bifurcations}.
For comparison with the the work of \cite{Gibson2009}, projections were only performed at three distinct Reynolds numbers of $400$, $330$ and $270$. The results are labelled accordingly.

\subsection{Equilibria catalogue}\label{sec:eq_props}
The properties of all previously known and new equilibria are given in table \ref{tab:equilibria}, 
where the solutions are organised by isotropy subgroup and sorted by descending dissipation rate, $D$, within.
In the following sections equilibria flow fields will be labelled as EQX rather than $\bs{u}_{EQX}$ for clarity.
The equilibria labels follow the naming convention of \cite{Gibson2009}; 
EQ0 is the laminar solution, 
EQ1 and EQ2 are \cite{Nagata1990}'s lower- and upper-branch equilibria, respectively, 
and EQ3-EQ11 are the equilibria discovered by \cite{Gibson2009} and \cite{Halcrow2008}.
The rest of the equilibria reported here are new and the numeric labelling denotes the chronological order in which they were found.
In the following, a projection of integer rank $M$ is denoted $\Pi_M(\cdot)$, where $1 \leq M \leq 99$.

\begin{table} % EQ table
	\centering
	\begin{tabular}{rlccccccccrr}
		\toprule
		\multicolumn{1}{c}{Root} & \multicolumn{1}{c}{$Re$}	& $\mathrm{EQ}$	& $\| \tilde{\bs{u}} \|$ 	& $E$ 		& $D$ 		& $H$ 			& dim($W^{u}$) 	& dim($W^{u}_{H}$) 	& Acc. 	& Ret. & Occ.\\ \midrule
		& 400	& mean			& 0.2997 			& 0.1016	& 2.6017 	& $\{e\}$ 		&				& 			& \\
		& 		& 0      		& 0.0000			& 0.1667 	& 1.0000	& $\Gamma$ 		& 0  			& 0  		& 				& 	  0	& 199	\\
		\rank{3}{4}		& 400	& 3      		& 0.1259    		& 0.1382	& 1.3177	& $\Sigma$ 		& 4  			& 2  		& $10^{-4}$		& 	 73	&  36 	\\
		& 400	& 4      		& 0.1681    		& 0.1243	& 1.4537	& $\Sigma$ 		& 6  			& 3  		& $10^{-6}$		& 	 73 &   2	\\ 
		\rank{1}{2}		& 400	& 1   			& 0.2091    		& 0.1363 	& 1.4293	& $\Sigma$ 		& 1  			& 1  		& $10^{-6}$		& 	 97 &  66	\\
		& 400	& 5      		& 0.2186   			& 0.1073	& 2.0201	& $\Sigma$ 		& 11 			& 4  		& $10^{-3}$		& 	 32	&   1	\\
		\rank{5}{23}	& 400	& 2      		& 0.3858    		& 0.0780	& 3.0437	& $\Sigma$ 		& 8  			& 2  		& $10^{-4}$		& 	 96	&  62	\\
		\rank{2}{5} 	& 400	& 7      		& 0.0936			& 0.1469	& 1.2523	& $\Theta$ 		& 3  			& 1  		& $10^{-4}$		&	 66 &  44	\\
		\rank{7}{4}		& 400	& 9      		& 0.1565    		& 0.1290	& 1.4048	& $\Theta_3$	& 5  			& 3  		& $10^{-4}$		& 	 74 &   6	\\
		& 400	& 10     		& 0.3285    		& 0.1080	& 2.3721	& $\Theta_3$	& 10 			& 7  		& $10^{-4}$		& 	 48 &   1	\\
		& 400	& 11     		& 0.4049 			& 0.0803	& 3.4322	& $\Theta_3$	& 13 			& 10 		& $10^{-3}$		& 	  1	&   1	\\ \midrule		
		\rank{5}{9}		& 400	& 20     		& 0.2405 			& 0.1289	& 1.6034	& $\Theta_3$	& 3  			& 2  		& $10^{-5}$		& 	 97 &  49	\\ 				% 16
		\rank{3}{10}	& 400	& 21     		& 0.2683 			& 0.1242	& 1.7630	& $\Theta_3$	& 4  			& 3  		& $10^{-6}$		& 	 43 &  79 	\\				% 17
		\rank{7}{10}	& 400	& 22     		& 0.3037 			& 0.1160	& 2.0713	& $\Theta_3$	& 8  			& 6  		& $10^{-5}$		& 	 40 &  25 	\\				% 18
		\rank{7}{11}	& 400	& 23     		& 0.4014 			& 0.0759	& 3.2474	& $\Theta_3$	& 10 			& 4  		& $10^{-5}$		& 	 33 &  40	\\ 				% 19
		\rank{10}{11}	& 400	& 24			& 0.4049			& 0.0813	& 3.3612	& $\Theta_3$	& 15			& 9			& $10^{-5}$		& 	  2 &  11 \\ 				% 28
		&		&				&					&			&			&				&				&			& 			  \\
		
		& 330	& mean			& 0.2541 			& 0.0959	& 1.6660 	& $\{e\}$ 		&				& 			& \\
		& 330 	& 6 			& 0.2751    		& 0.0972	& 2.8185	& $\Sigma$ 		& 19 			& 6  		& $10^{-3}$		&  	16 &  6	\\ \midrule
		\rank{6}{6}		& 330 	& 13  			& 0.2168 			& 0.1337	& 1.4705	& $\Sigma$ 		& 1  			& 1  		& $10^{-3}$		& 	98 & 26	\\  			% 13
		\rank{40}{6}	& 330 	& 14 			& 0.2375 			& 0.1052	& 2.2785	& $\Sigma$		& 15 			& 6  		& $10^{-2}$		& 	57 & 42 \\  			% 12
		\rank{2}{6}		& 330 	& 12 			& 0.1145 			& 0.1410	& 1.3433	& $\Theta$ 		& 3  			& 1  		& $10^{-3}$		& 	67 & 92 \\    			% 15
		\rank{67}{6}	& 330	& 16			& 0.2348			& 0.1063	& 2.3047	& $\Theta_3$	& 15			& 8			& $10^{-2}$		& 	17 &  5 \\  			% 29
		\rank{42}{6}	& 330 	& 15  			& 0.2674 			& 0.0988	& 2.6947	& $\Theta_3$ 	& 18  			& 9  		& $10^{-2}$		& 	27 & 37 \\ \midrule  	% 24
		\rank{31}{15}	& 330	& 25 	 		& 0.2331			& 0.1292	& 1.5650	& $\Theta_3$	& 3	 			& 2  		& $10^{-3}$		& 	93 &  7 \\  			% 33
		\rank{66}{15}	& 330	& 26 	 		& 0.2707			& 0.0975	& 2.6274	& $\Theta_3$	& 17			& 9  		& $10^{-2}$		& 	40 &  1 \\  			% 26
		&		&				&					&			&			&				&				&			& 			  \\
		
		& 270	& mean			& 0.2286 			& 0.1089	& 1.4813 	& $\{e\}$ 		&				& 			& \\
		& 270 	& 8 			& 0.3466 			& 0.0853	& 3.6719	& $\Theta$ 		& 15 			& 2  		& $10^{-3}$		& 	42 &  1 \\ \midrule
		\rank{11}{8}	& 270	& 18 	 		& 0.2292			& 0.1294	& 1.5415	& $\Sigma$		& 1	 			& 1  		& $10^{-2}$		& 	34 &  9 \\  			% 25
		\rank{3}{8}		& 270 	& 17 			& 0.1546 			& 0.1301	& 1.5530	& $\Theta$ 		& 5  			& 1  		& $10^{-2}$		& 	67 & 32 \\  			% 14
		\rank{38}{8}	& 270 	& 19  			& 0.3148 			& 0.0904	& 3.1529	& $K$ 			& 12 			& 0  		& $10^{-2}$		& 	67 &  9 \\ \midrule  	% 20
		\rank{2}{18}	& 270	& 27 	 		& 0.2297			& 0.1292	& 1.5444	& $\Theta_3$	& 2				& 2  		& $10^{-2}$		& 	91 & 64 \\ \bottomrule 	% 27
	\end{tabular}
	\caption{Properties of the new equilibria and the equilibria they are derived from; 
		a dividing line separates parent equilibria (previously known solutions) from their child equilibria (new solutions) from their grandchild equilibria (solutions found from projections of the child equilibria).
		The `Root' column denotes the initial velocity field that lead to the new discovery, e.g. a rank-5 projection of $\mathrm{EQ}9$ lead to the discovery of $\mathrm{EQ}16$.
		The `$Re$' column indicates the distinct Reynolds number that the solution was discovered at.
		The `mean' values are given for comparison, and are calculated from a spatially and temporally averaged turbulent flow.
		The $\mathscr{L}^{2}$-norm of the velocity field is $\| \tilde{\bs{u}} \|$,
		$E$ is the kinetic energy density,
		$D$ is the dissipation rate,
		$H$ is the isotropy subgroup,
		dim($W^{u}$) is the dimensionality of the equilibrium's unstable manifold,
		and dim($W^{u}_{H}$) is the dimensionality of the unstable manifold within the $H$-invariant subspace.
		The accuracy of the solution (Acc.) is calculated using \eqnref{eqn:acc}.
		The return column (Ret.) denotes how often the solution was found using projections of itself.
		The occurrence column (Occ.) denotes how often the solution was found from searches initiated using projections of other equilibria.}
	\label{tab:equilibria}
\end{table}

\begin{figure} % Positive real unstable eigenvalues for all Reynolds numbers
	\begin{subfigure}{\textwidth}
		\centering
		\includegraphics[width=.9\linewidth]{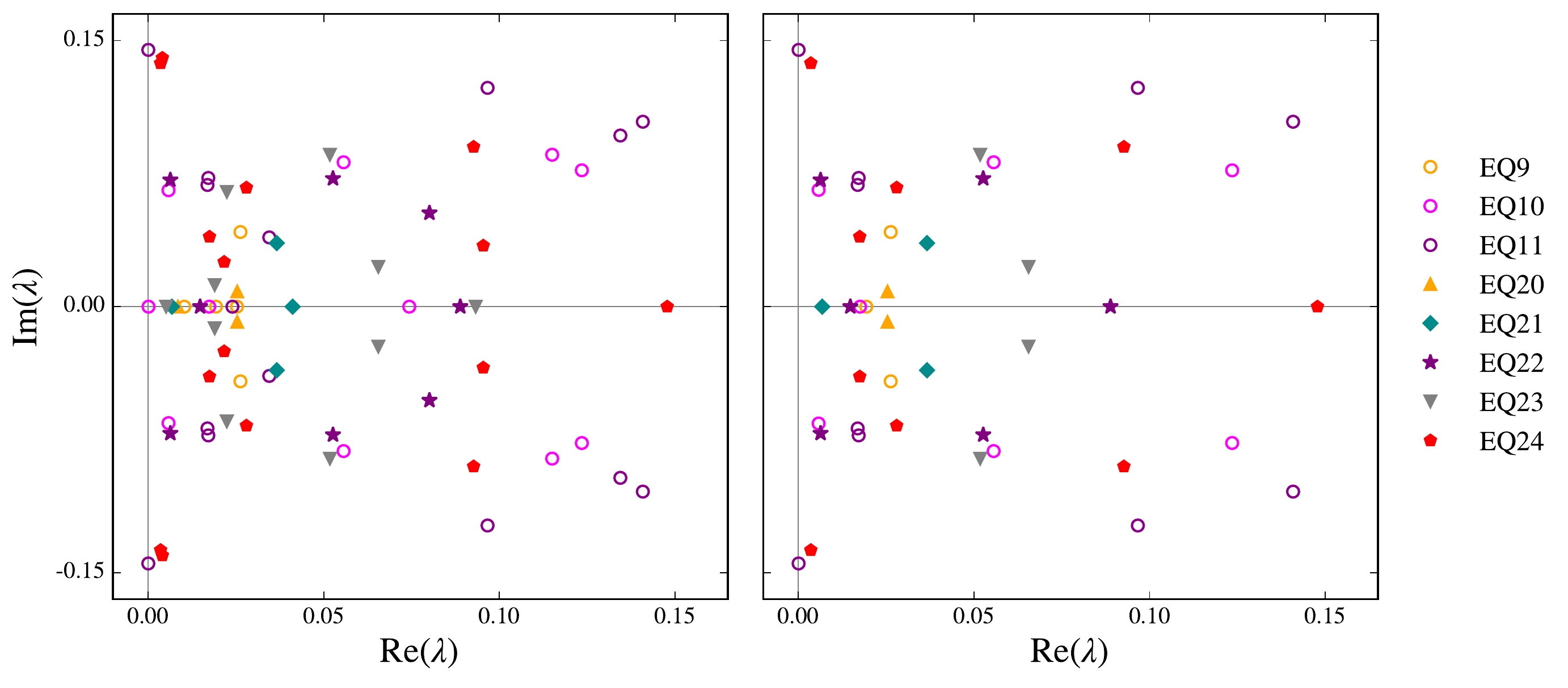}\vspace*{-0.2cm}
		\caption{}
		\label{fig:re400-eigs}
	\end{subfigure}
	\vspace{0.2cm}
	\begin{subfigure}{\linewidth}
		\centering
		\includegraphics[width=.9\linewidth]{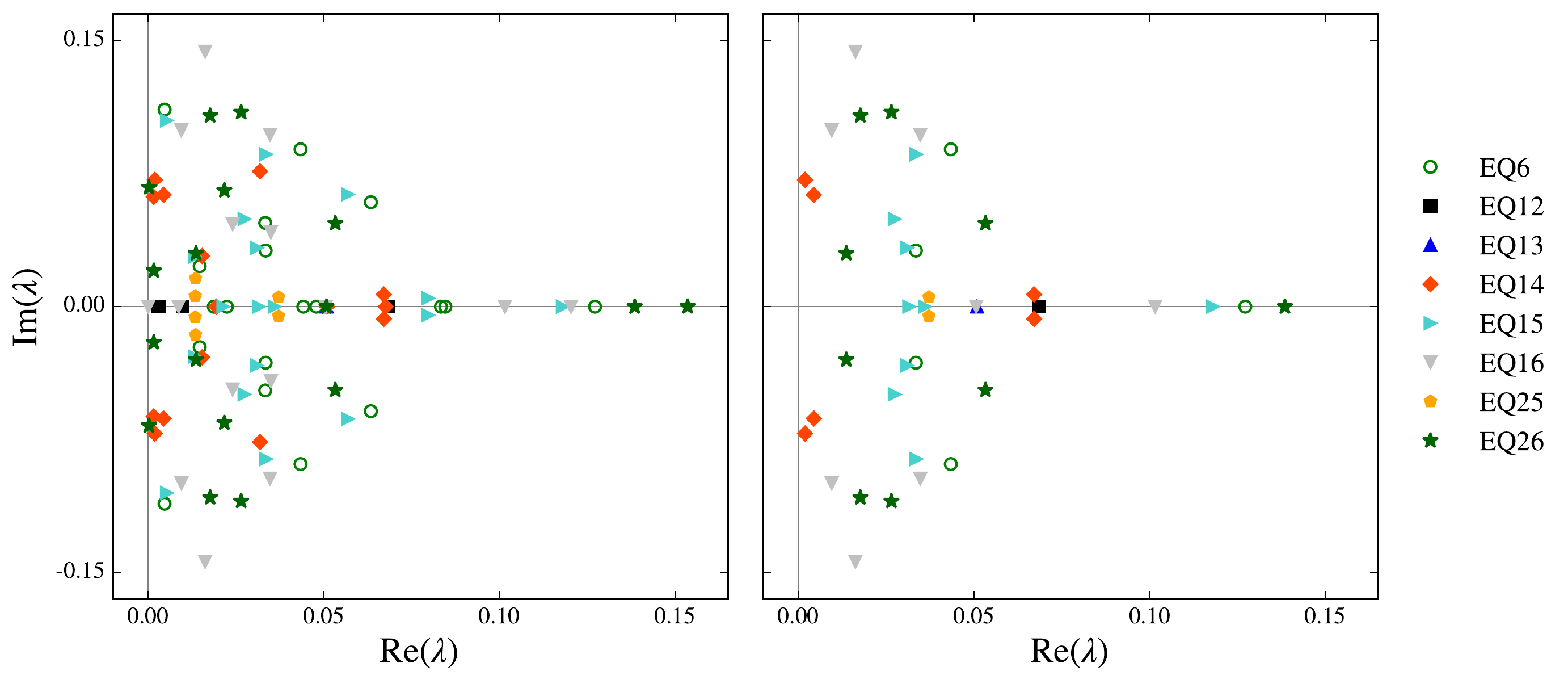}\vspace*{-0.2cm}
		\caption{}
		\label{fig:re330-eigs}
	\end{subfigure}
	\vspace{0.2cm}
	\begin{subfigure}{\linewidth}
		\centering
		\includegraphics[width=.9\linewidth]{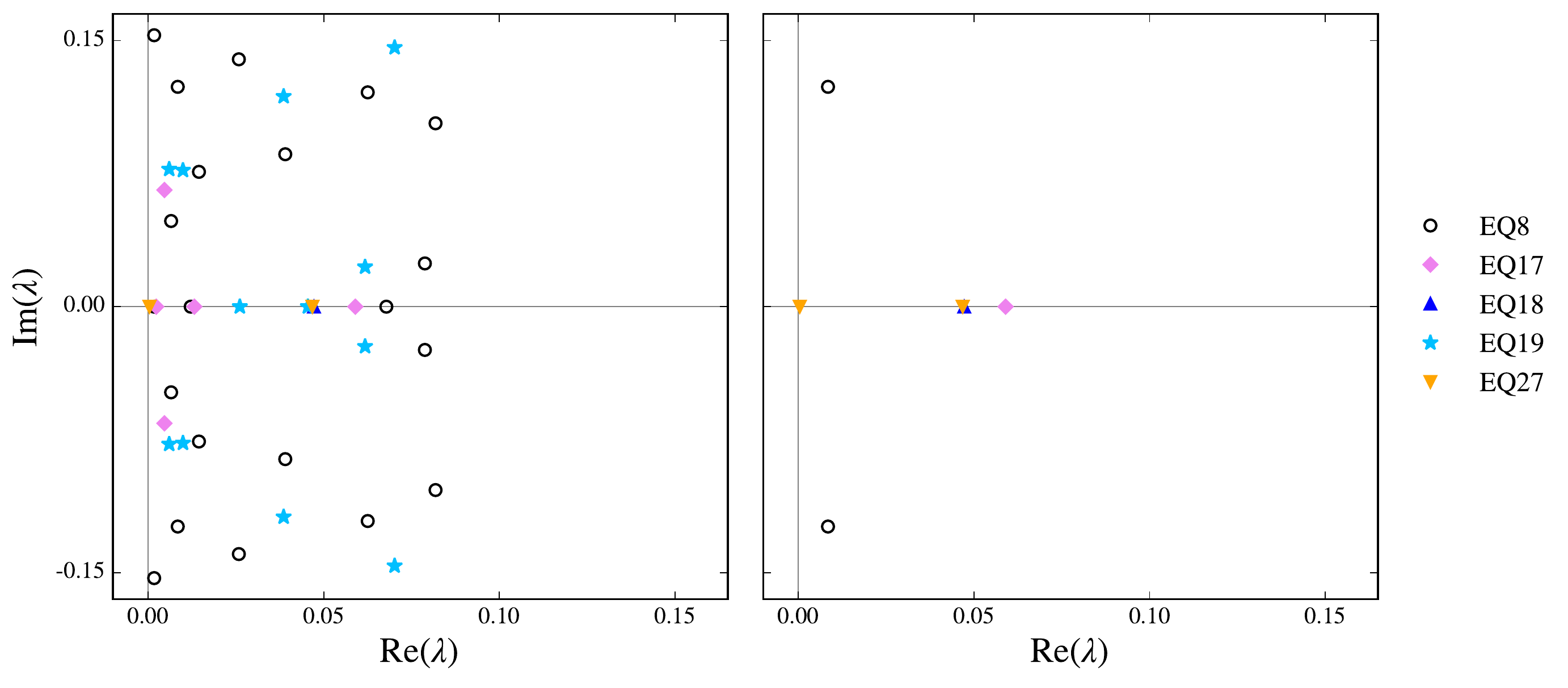}\vspace*{-0.2cm}
		\caption{}
		\label{fig:re270-eigs}
	\end{subfigure}
	\caption{Unstable eigenvalues ($\lambda$) of equilibria and their child equilibria in the full space $\mathbb{U}$ (left), and symmetry-invariant subspace $\mathbb{U}_{H}$ (right), where $H$ is a place-holder for the symmetry subgroup of the equilibria. These were calculated for: (a) $Re = 400$, (b) $Re = 330$ and (c) $Re = 270$.}
	\label{fig:eigs}
\end{figure}

% What is shown in the eigenvalues figure?
The unstable parts of the new equilibria and their parents' spectra are shown in figure \ref{fig:eigs}.
From this figure we can get an idea of the nature of their unstable manifolds computed using the leading unstable eigenvalues and eigenvectors.
For example, the leading unstable eigenvalues of EQ11 are a complex conjugate pair, 
so we expect the trajectories, perturbed along the leading unstable eigenvector of EQ11, to travel on a surface away from the fixed point.
Whereas, for EQ24 the leading unstable eigenvalue is real only, hence we expect all perturbed trajectories to travel along a one-dimensional curve.
For figures depicting these unstable manifolds, see \S\ref{sec:eq_state_space}.

\subsubsection{Child equilibria found at $Re = 400$}
\paragraph{\textbf{EQ20, EQ21, EQ23.}} % EQ16 17 19
These equilibria were discovered from projections of EQ9 and EQ10.
The new solutions obey the same symmetries as the parent equilibria and all three of them resemble EQ10; the fields contain a streamwise streak flanked by a pair of counter-rotating asymmetric vortices.
The main difference between the flow fields is that the child equilibria have smoother structures with lower velocity magnitudes. 
The unstable manifolds of the child solutions have reduced dimensionality in $\mathbb{U}$ and within the $\Theta_3$-invariant subspace $\mathbb{U}_{\Theta3}$, as compared to the parent solutions.
The magnitudes of the unstable eigenvalues of the children are reduced compared to those of their parents, see figure \ref{fig:re400-eigs}.

\paragraph{\textbf{EQ22, EQ24}} % EQ18 28
These equilibria were discovered from projections of EQ11.
These equilibria were also found by \cite{Gibson2009} by continuation of EQ10 with varying spanwise wavenumber.
EQ22 resembles EQ10 and EQ24 is similar in appearance to EQ11 
and both children and parents adhere to the $\Theta_3$ isotropy subgroup.
Both EQ22 and EQ24 have reduced dimensionality in $\mathbb{U}_{\Theta3}$.
However, EQ24 is more unstable than EQ11 in $\mathbb{U}$ and further from the laminar state, so presumably it is more likely to be approached by a turbulent trajectory.
The unstable eigenvalues of the new equilibria and EQ11 have similar real parts in both $\mathbb{U}$ and $\mathbb{U}_{\Theta3}$, see figure \ref{fig:re400-eigs}.

\subsubsection{Child equilibria found at $Re = 330$}
All of the following equilibria were discovered from projections of EQ6.

\paragraph{$\bs{\mathrm{EQ12}}$.} % EQ15
This solution sits on the lower branch of the EQ7-8 bifurcation curve at $Re=330$, see figure \ref{fig:bifs_all}.
The solution looks like EQ7 in terms of velocity field structure and also obeys the same symmetries as the solutions on the EQ7-8 branch (all solutions on the curve belong adhere to the $\Theta$-isotropy subgroup).
Of all the equilibria in table \ref{tab:equilibria}, it has one of the lowest energy norms and dissipation rates, both properties being close to the laminar solution.
This is not surprising as the solution is on the EQ7 branch, which is the closest solution to the laminar state.

\paragraph{$\bs{\mathrm{EQ13}}$.} % EQ13
This solution sits on \citeauthor{Nagata1990}'s lower branch at $Re=330$ (see figure \ref{fig:bifs_400}), accordingly it resembles and has the same symmetries as EQ1.
Similar to EQ1, the new solution has a one dimensional unstable manifold within in the $\Sigma$-invariant subspace and the full space, and the magnitude of their unstable eigenvalues are similar.

\paragraph{$\bs{\mathrm{EQ14}}$.} % EQ12
This equilibrium solution sits on the lower branch of the EQ5-6 bifurcation curve at $Re=330$, see figure \ref{fig:bifs_330}. 
EQ14 forms a lower-upper branch pair with EQ6 at $Re = 330$, consequently the solution looks similar to and belongs to the same isotropy subgroup as its upper-branch partner.
EQ14's unstable manifold has the same dimensionality in the $\Sigma$-invariant subspace as EQ6.

\paragraph{$\bs{\mathrm{EQ16}}, \bs{\mathrm{EQ26}}$.} % EQ29 26
These solutions are located on a new branch shown in figure \ref{fig:bifs_all}.
The flow field structures of the new solutions are similar to EQ6 with two fast-travelling streamwise streaks and two slow-travelling streamwise streaks above, arranged in an alternating fashion in the spanwise direction.
The child equilibria belong to a different isotropy subgroup than their parent solution;
the structures within EQ16 and EQ26 are asymmetric with respect to rotation and reflection symmetries, hence they belong to $\Theta_3$.
Both solutions have high dimensional unstable manifolds in $\mathbb{U}_{\Theta3}$, 
and their unstable manifolds have even greater dimensionality than their parent in $\mathbb{U}$.
The magnitudes of the unstable eigenvalues are also higher than other solutions discovered at $Re=330$, see figure \ref{fig:re330-eigs}.

\paragraph{$\bs{\mathrm{EQ15}}$.} % EQ24
This equilibrium solution inhabits a newly discovered branch.
It is similar to EQ6 in terms of flow field structures, however its symmetry subgroup is different; 
the solution is part of the $\Theta_3$-isotropy subgroup, hence it is less symmetric than its parent.
EQ15 is highly dissipative and very unstable; its unstable manifold has greater dimensionality than EQ6 in the full space and the symmetry-invariant subspace.
It is the most unstable solution found at $Re=330$, both its energy norms and dissipation rates are close to those of EQ6.

\paragraph{$\bs{\mathrm{EQ25}}$.} % EQ33
This solution sits on the lower branch of the EQ20-21-23 curve at $Re = 330$.
In terms of field structure the solution is identical to EQ20, except it is reflected in the $yx$-plane, i.e. EQ25 $= \sigma_z$EQ20.
It also has the same number of unstable eigenvalues as EQ20 in the full space and symmetry-invariant subspace, 
and also belongs to the same isotropy subgroup as EQ20.

\subsubsection{Child equilibria found at $Re = 270$}\label{sec:EQ@270}
All of the following equilibria were discovered from projections of EQ8.

\paragraph{$\bs{\mathrm{EQ17}}$.} % EQ14
This solution sits on the lower branch of the EQ7-8 bifurcation curve at $Re = 270$ and shares much of its properties with EQ7.
The solution looks like EQ7 in terms of velocity field structure and obeys the same symmetries as all solutions on the EQ7-8 bifurcation curve.

\paragraph{$\bs{\mathrm{EQ18}}$.} % EQ25
This solution is situated on the lower branch of EQ1-2 bifurcation curve at $Re = 270$, it has the same symmetries as and looks similar to EQ1.
The new equilibrium also has the same number and similar magnitude of unstable eigenvalues as EQ1.

\paragraph{$\bs{\mathrm{EQ19}}$.} % EQ20
This solution is located on a new branch shown in figure \ref{fig:bifs_all}.
It looks like an asymmetric version of EQ2 and is very dissipative.
The solution does not have an unstable manifold in the symmetry-invariant subspace, 
which has not been seen before for any type of solution, 
and in the full space it has many unstable eigenvalues and hence a high dimensional unstable manifold.
EQ19 is part of a symmetry subgroup previous solutions have not been observed to belong to. 
It is labelled as $K$ in table \ref{tab:equilibria}, defined as $K= \{ e, \theta_1, \theta_2, \theta_3\}$ (see Section \ref{sec:Symms} for more details).
The isotropy subgroup $K$ is unlike other isotropy subgroups as it allows the shift-reflect and shift-rotate symmetries given in \eqnref{eqn:symms} but only permits translations in the spanwise direction.

\paragraph{$\bs{\mathrm{EQ27}}$.} % EQ27
This solution, like EQ18 sits on the lower branch of the EQ1-2 bifurcation curve, however when continued in $Re$ it follows the path of the EQ20-21-23 bifurcation curve and is very similar to EQ25. 
Like EQ18, this is an interesting solution being close to the bifurcation points of many difference branches in figure \ref{fig:bifs_all}.

\subsection{Visualising equilibria in state space}\label{sec:eq_state_space}
\begin{figure} % Equilibria in state space Re = 400
	\centering
	\begin{minipage}{\textwidth}
		\centering
		\begin{subfigure}{0.55\linewidth}
			\centering
			\includegraphics[scale=0.47]{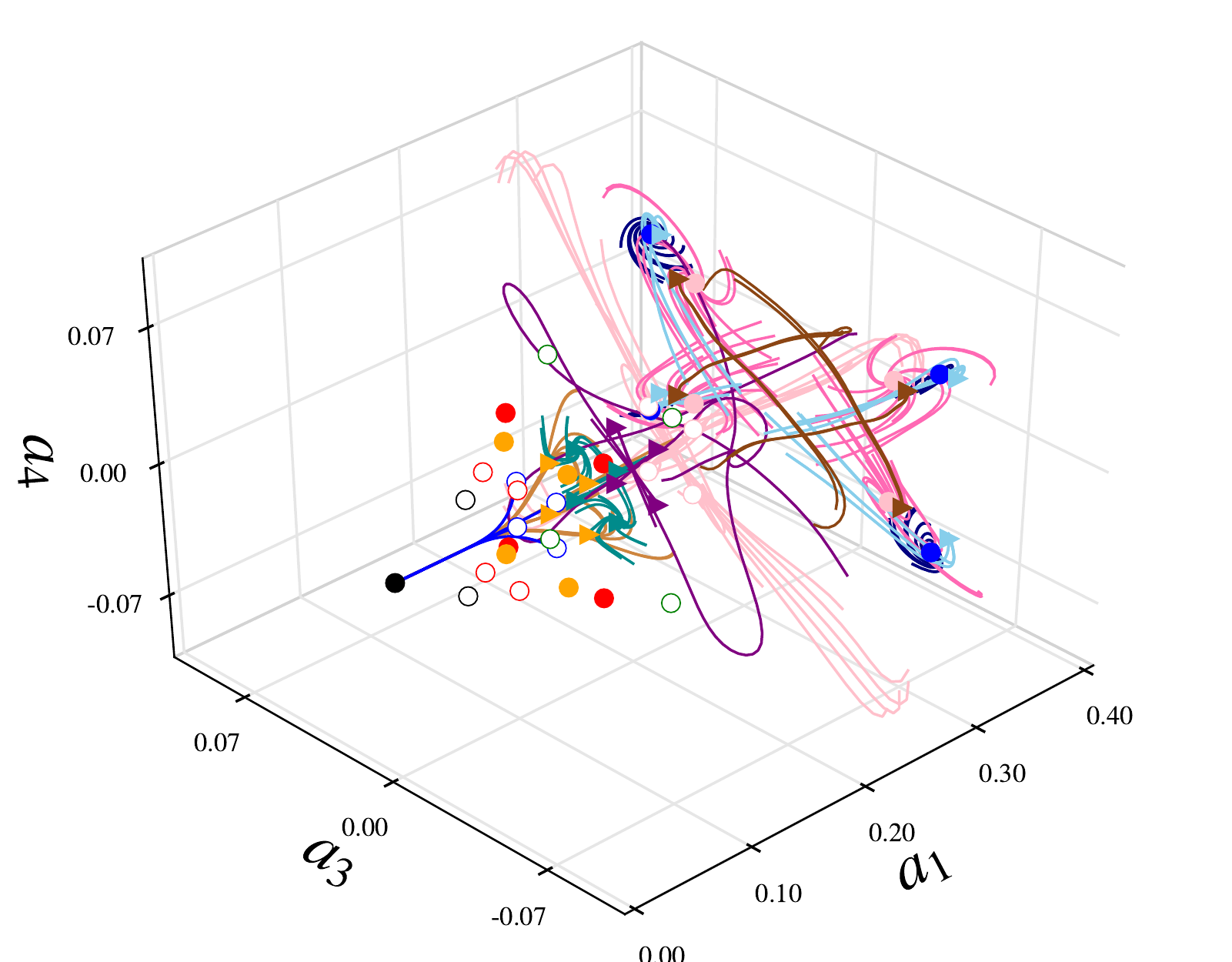}
			\caption{}
			\label{fig:re400_state_space_a}
		\end{subfigure}
		\begin{subfigure}{0.42\linewidth}
			\centering
			\includegraphics[width=\linewidth]{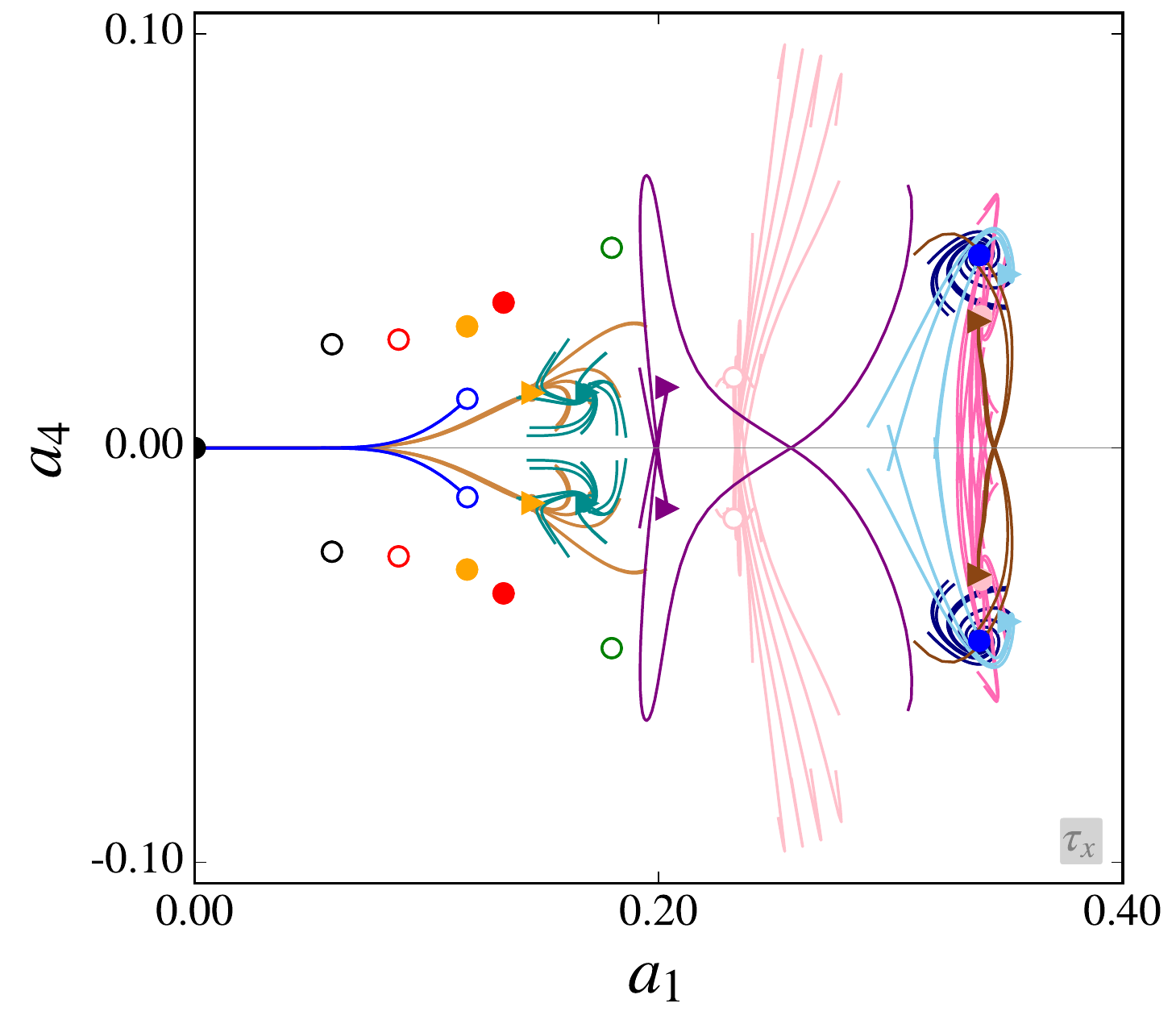}
			\caption{}
			\label{fig:re400_state_space_b}
		\end{subfigure}
	\end{minipage}\\%[-0.5ex]
	\begin{minipage}{\textwidth}
		\centering
		\begin{subfigure}{0.55\linewidth}
			\centering
			\includegraphics[scale=0.47]{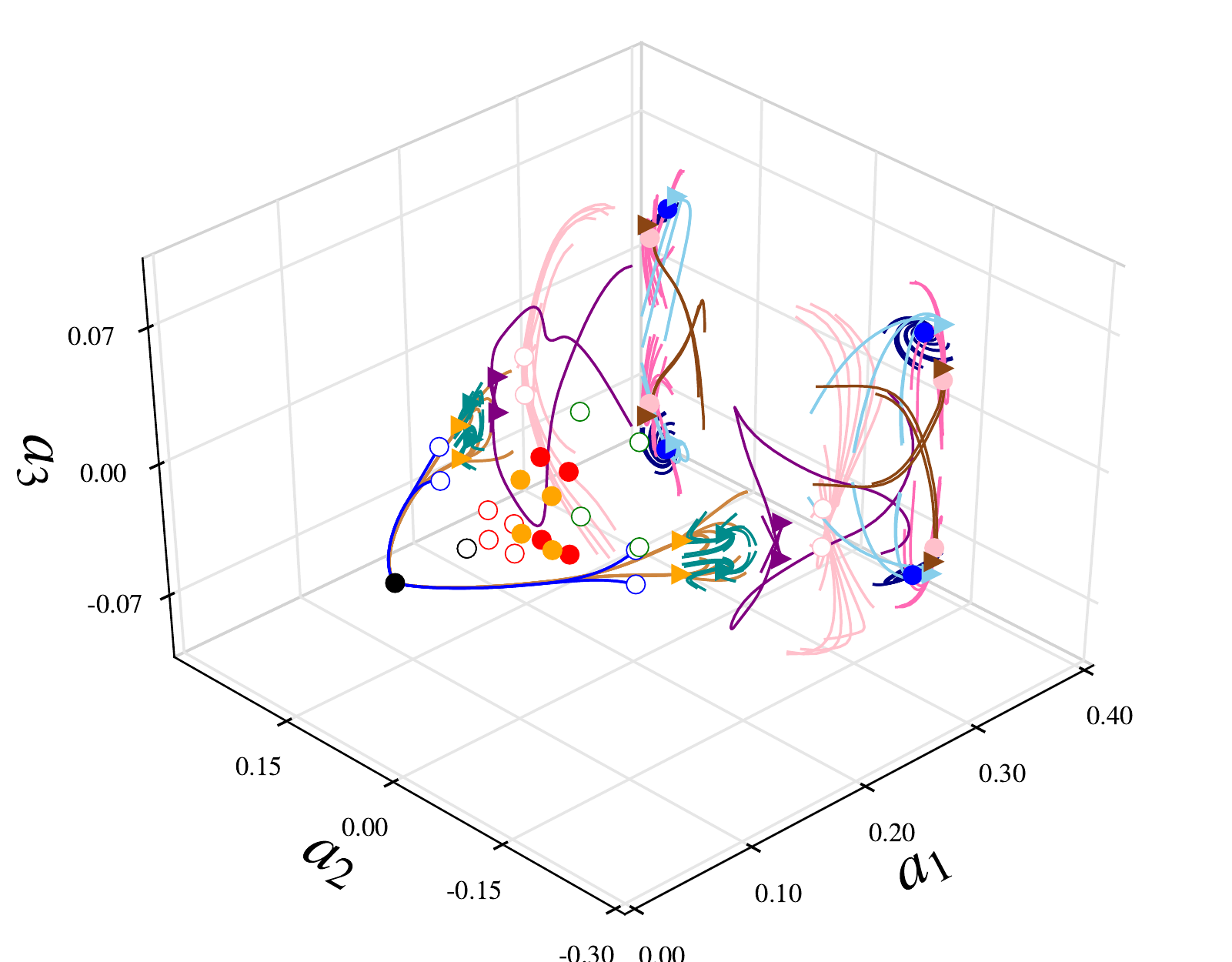}
			\caption{}
			\label{fig:re400_state_space_c}
		\end{subfigure}%\\[-1ex]
		\begin{subfigure}{0.42\linewidth}
			\centering
			\includegraphics[width=\linewidth]{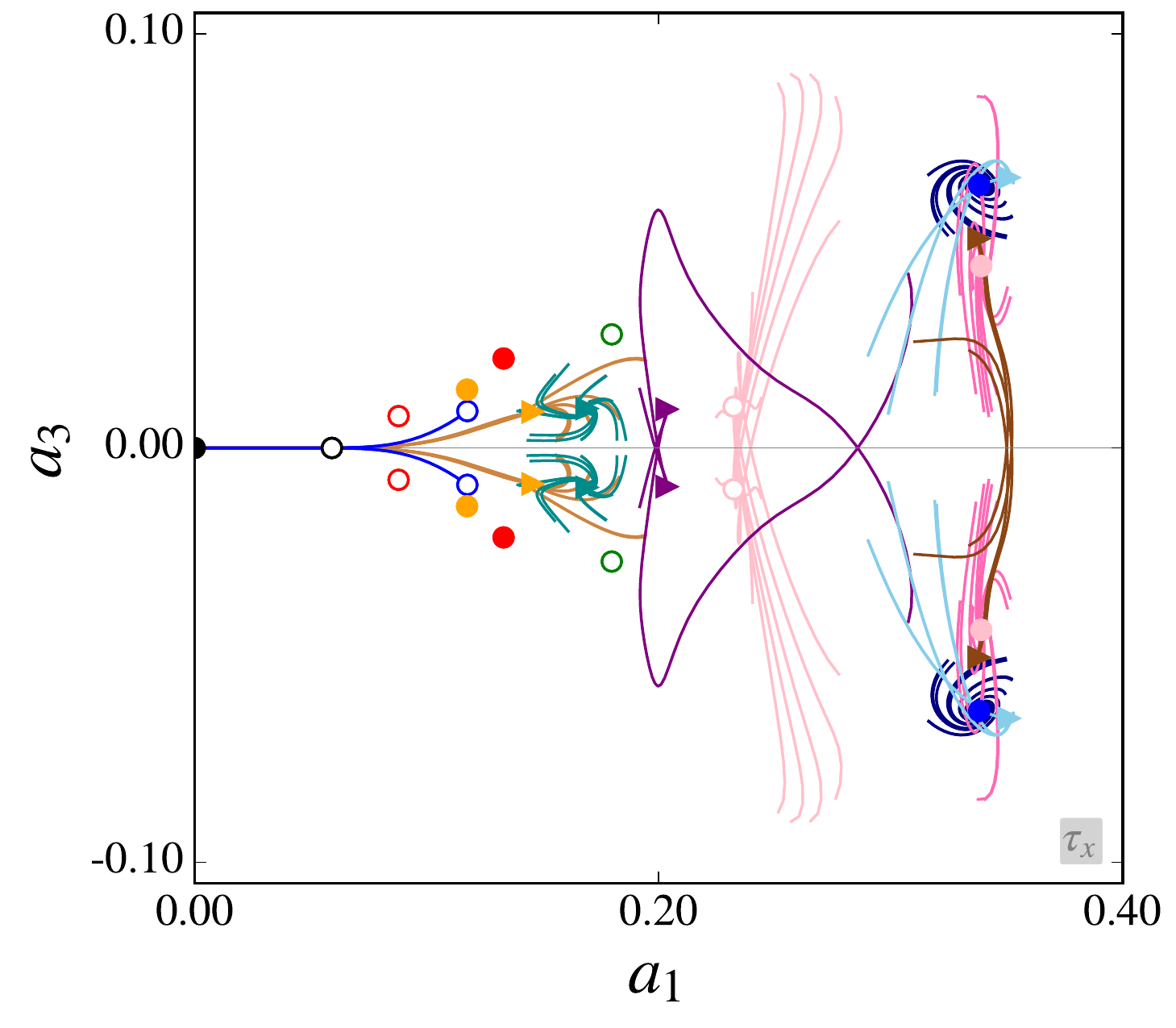}
			\caption{}
			\label{fig:re400_state_space_d}
		\end{subfigure}
	\end{minipage}
	\begin{minipage}{.65\textwidth}
		\centering
		\begin{subfigure}{\linewidth}
			\centering
			\includegraphics[width=\linewidth]{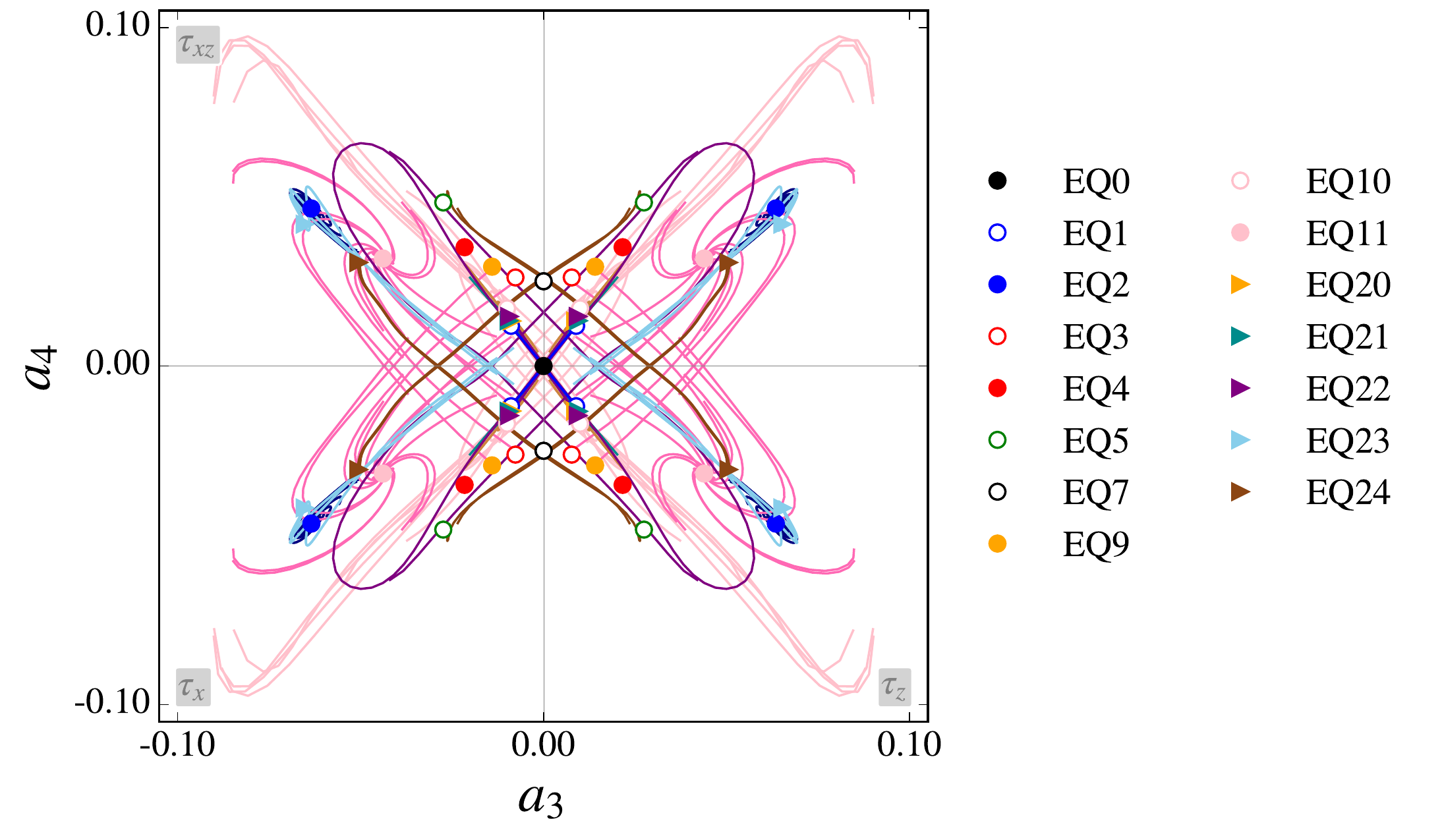}
			\caption{}
			\label{fig:re400_state_space_e}
		\end{subfigure}
	\end{minipage}
	\caption{Equilibria discovered at $Re=400$ in state space. The previously known equilibria are plotted with circular points and the new equilibria are plotted with triangular markers.
	The trajectories emanating from the markers represent fields that have been perturbed in the positive and negative eigendirections in state space.}
	\label{fig:re400_state_space}
\end{figure}

\begin{figure} % Equilibria in state space Re = 330
	\centering
	\begin{minipage}{\textwidth}
		\centering
		\begin{subfigure}{0.55\linewidth}
			\centering
			\includegraphics[scale=0.47]{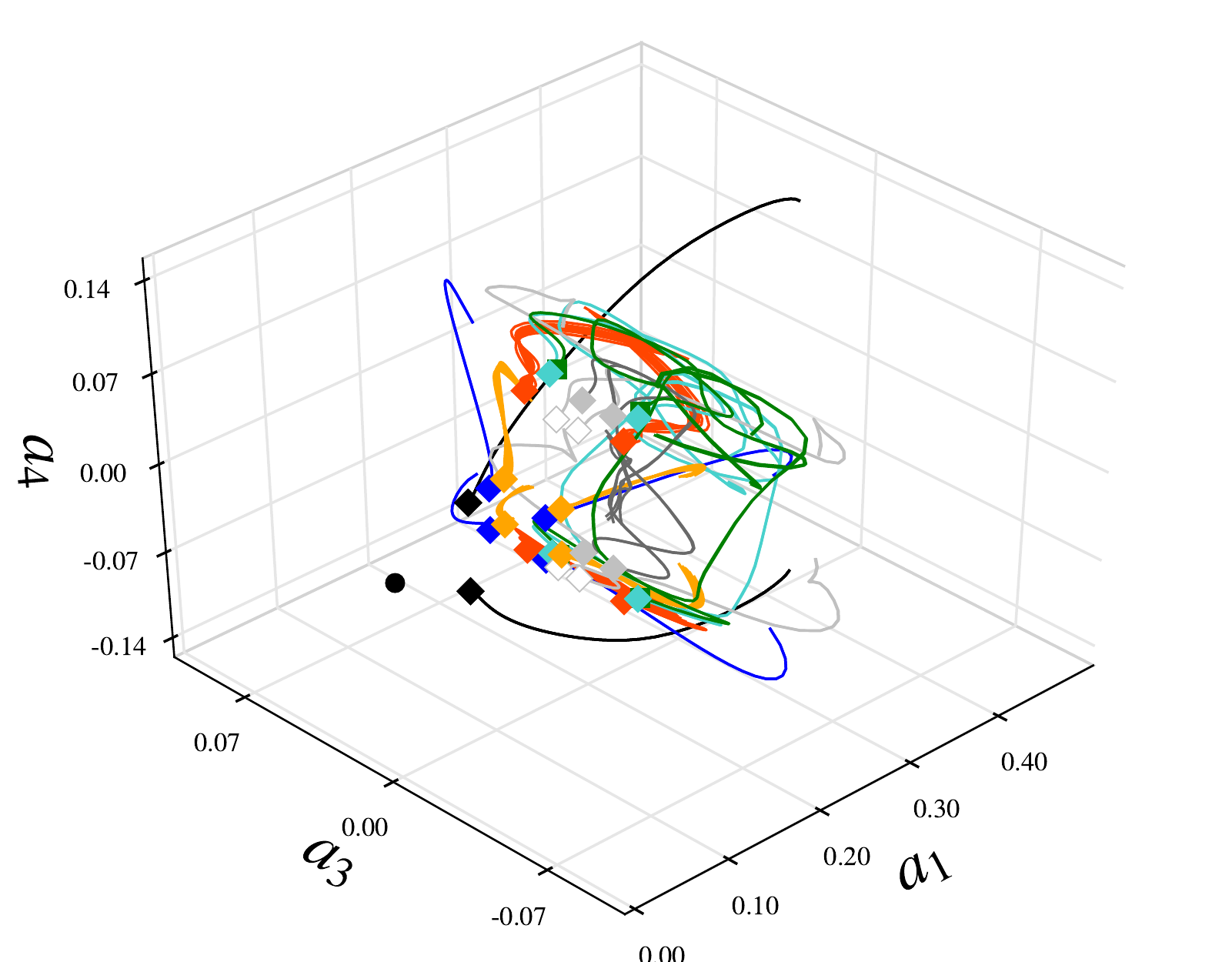}
			\caption{}
			\label{fig:re330_state_space_a}
		\end{subfigure}
		\begin{subfigure}{0.42\linewidth}
			\centering
			\includegraphics[width=\linewidth]{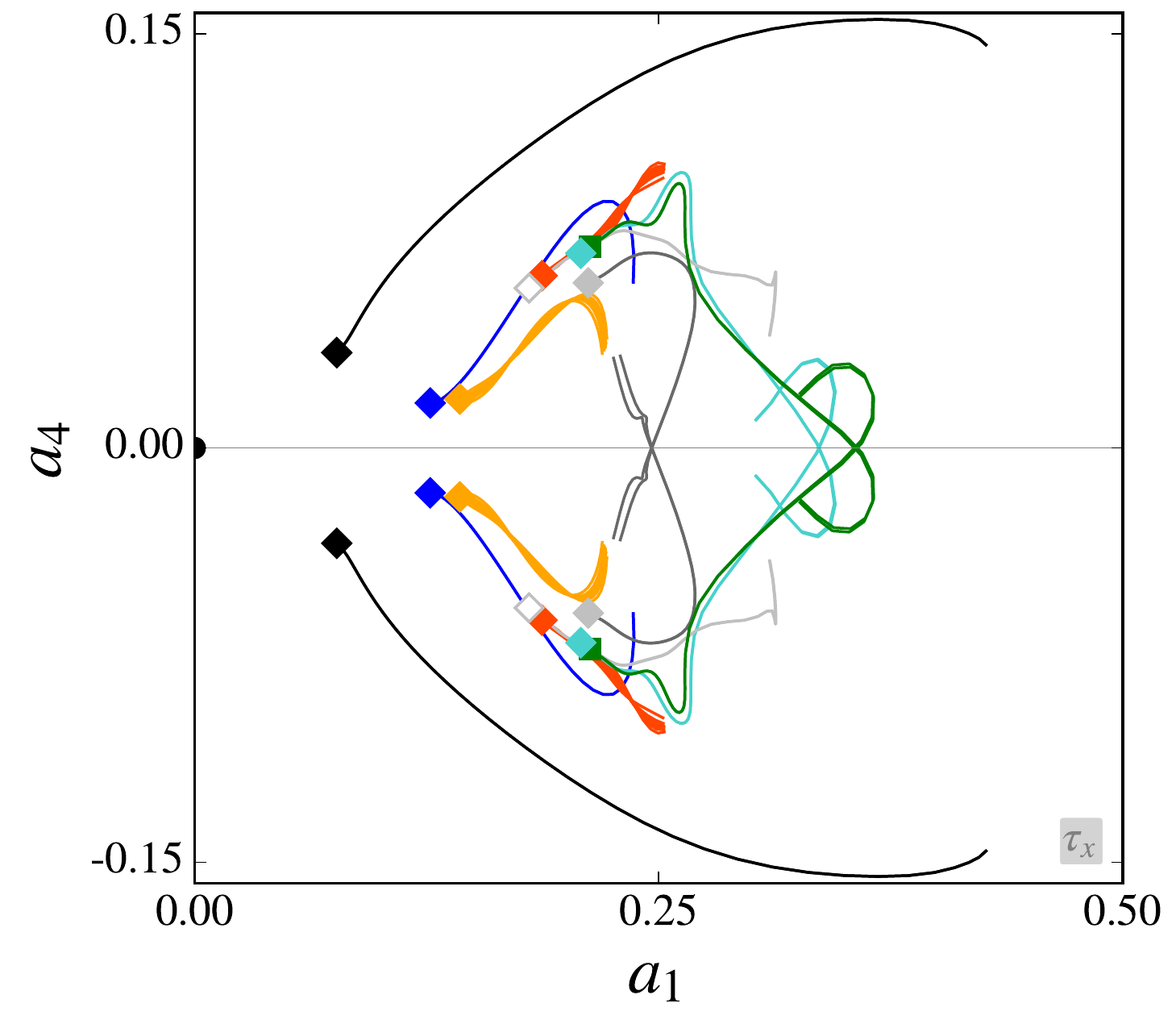}
			\caption{}
			\label{fig:re330_state_space_b}
		\end{subfigure}
	\end{minipage}\\
	\begin{minipage}{\textwidth}
		\centering
		\begin{subfigure}{0.55\linewidth}
			\centering
			\includegraphics[scale=0.47]{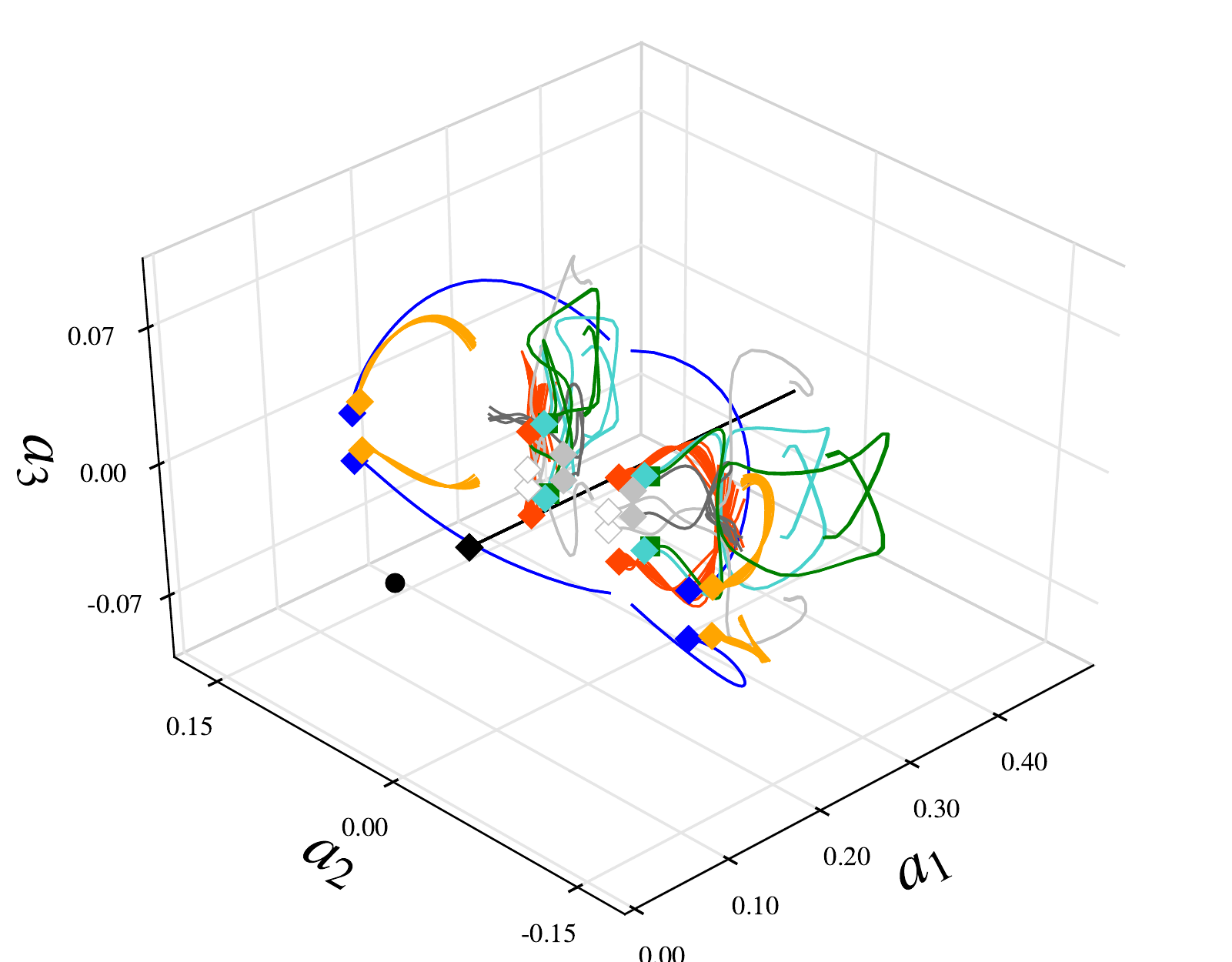}
			\caption{}
		\end{subfigure}
		\begin{subfigure}{0.42\linewidth}
			\centering
			\includegraphics[width=\linewidth]{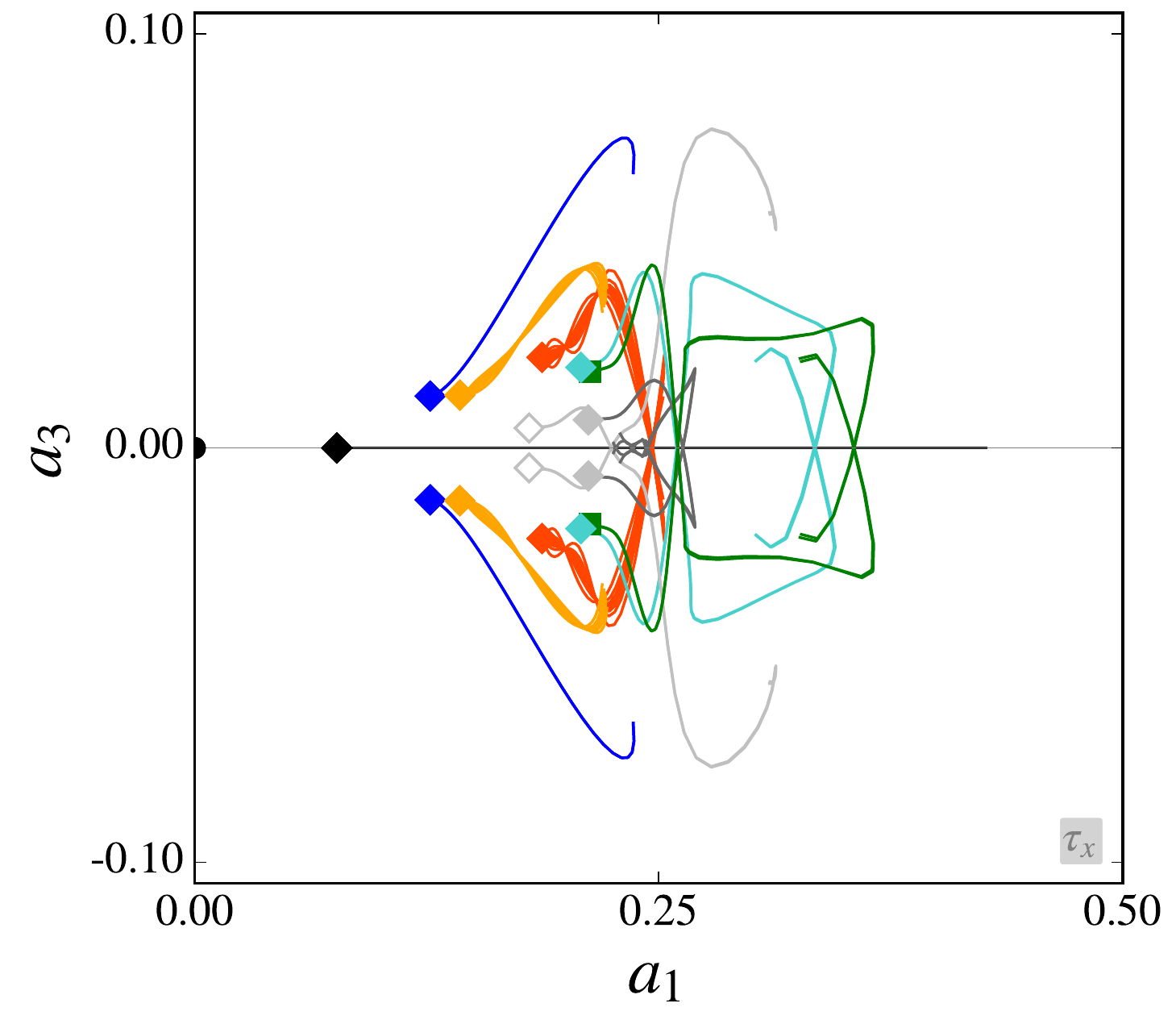}
			\caption{}
		\end{subfigure}
	\end{minipage}
	\begin{minipage}{.65\textwidth}
		\centering
		\begin{subfigure}{\linewidth}
			\centering
			\includegraphics[width=\linewidth]{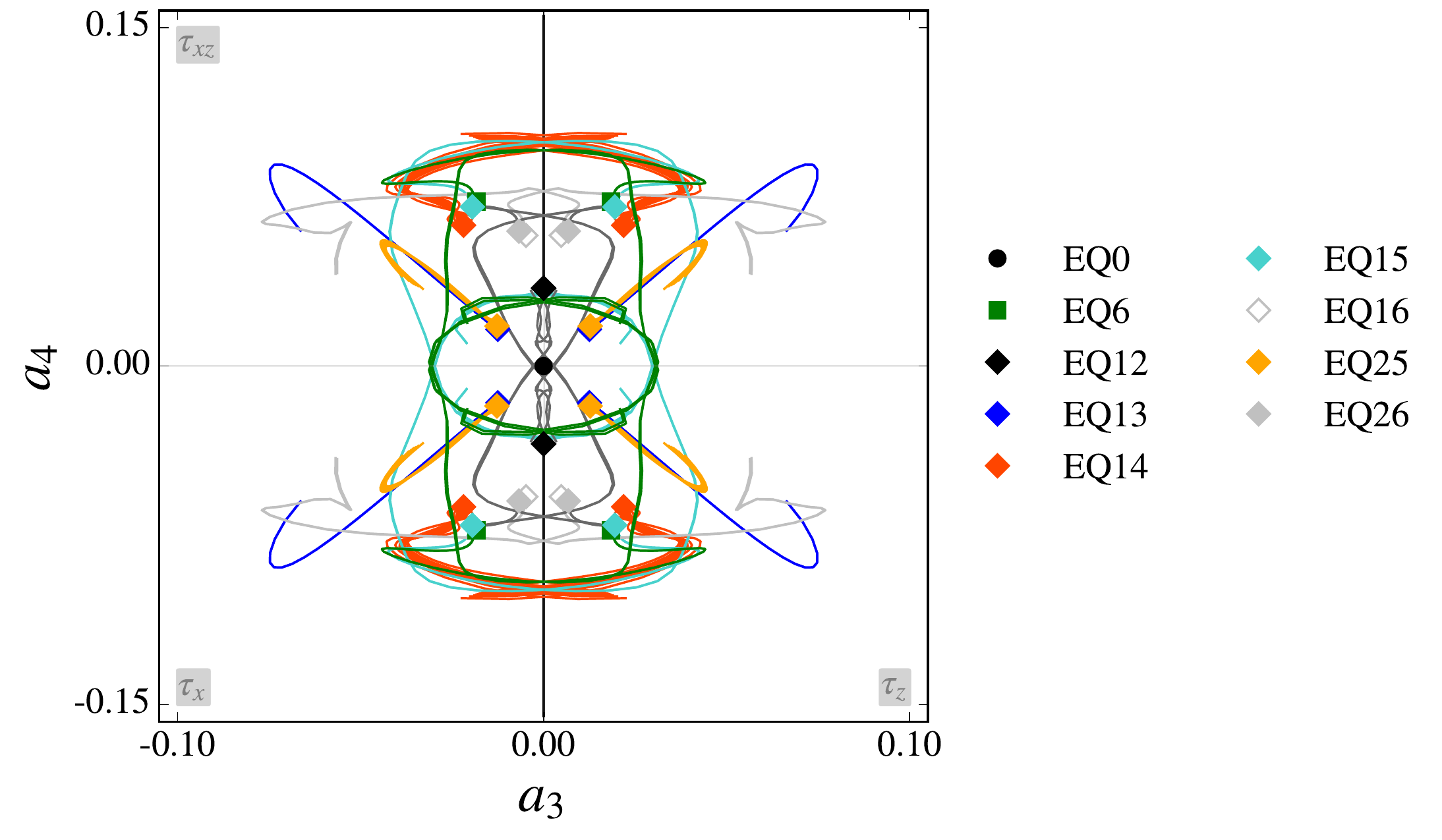}
			\caption{}
		\end{subfigure}
	\end{minipage}
	\caption{Equilibria discovered at $Re=330$ in state space. The previously known equilibrium solution (EQ6) is plotted with a square marker and the new equilibria are plotted with diamond markers. 
	The trajectories emanating from the markers represent fields that have been perturbed in the positive and negative eigendirections in state space.}
	\label{fig:re330_state_space}
\end{figure}

\begin{figure} % Equilibria in state space Re = 270
	\centering
	\begin{minipage}{\textwidth}
		\centering
		\begin{subfigure}{0.55\linewidth}
			\centering
			\includegraphics[scale=0.47]{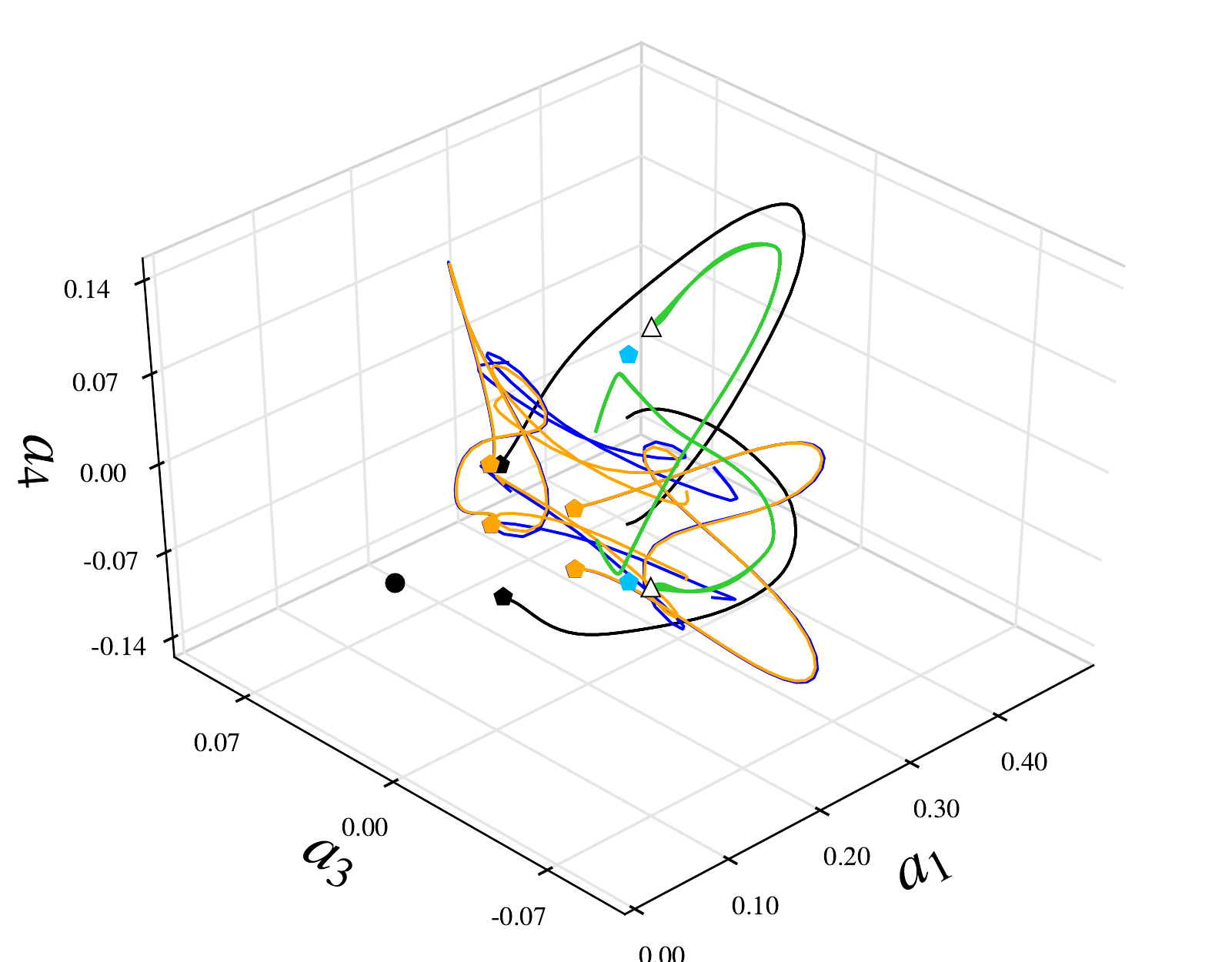}
			\caption{}
			\label{fig:re270_state_space_a}
		\end{subfigure}
		\begin{subfigure}{0.42\linewidth}
			\centering
			\includegraphics[width=\linewidth]{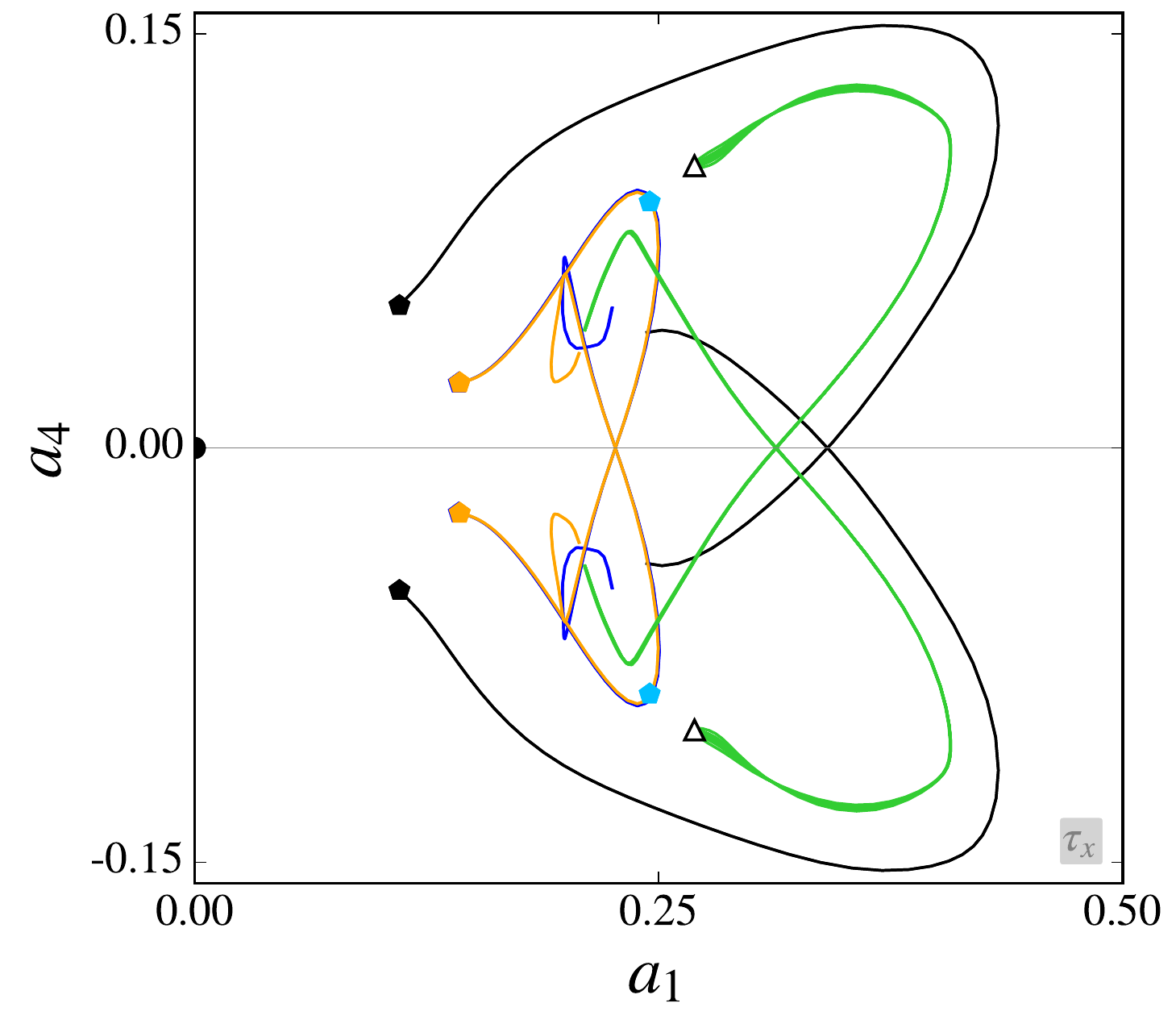}
			\caption{}
			\label{fig:re270_state_space_b}
		\end{subfigure}
	\end{minipage}\\
	\begin{minipage}{\textwidth}
		\centering
		\begin{subfigure}{0.55\linewidth}
			\centering
			\includegraphics[scale=0.47]{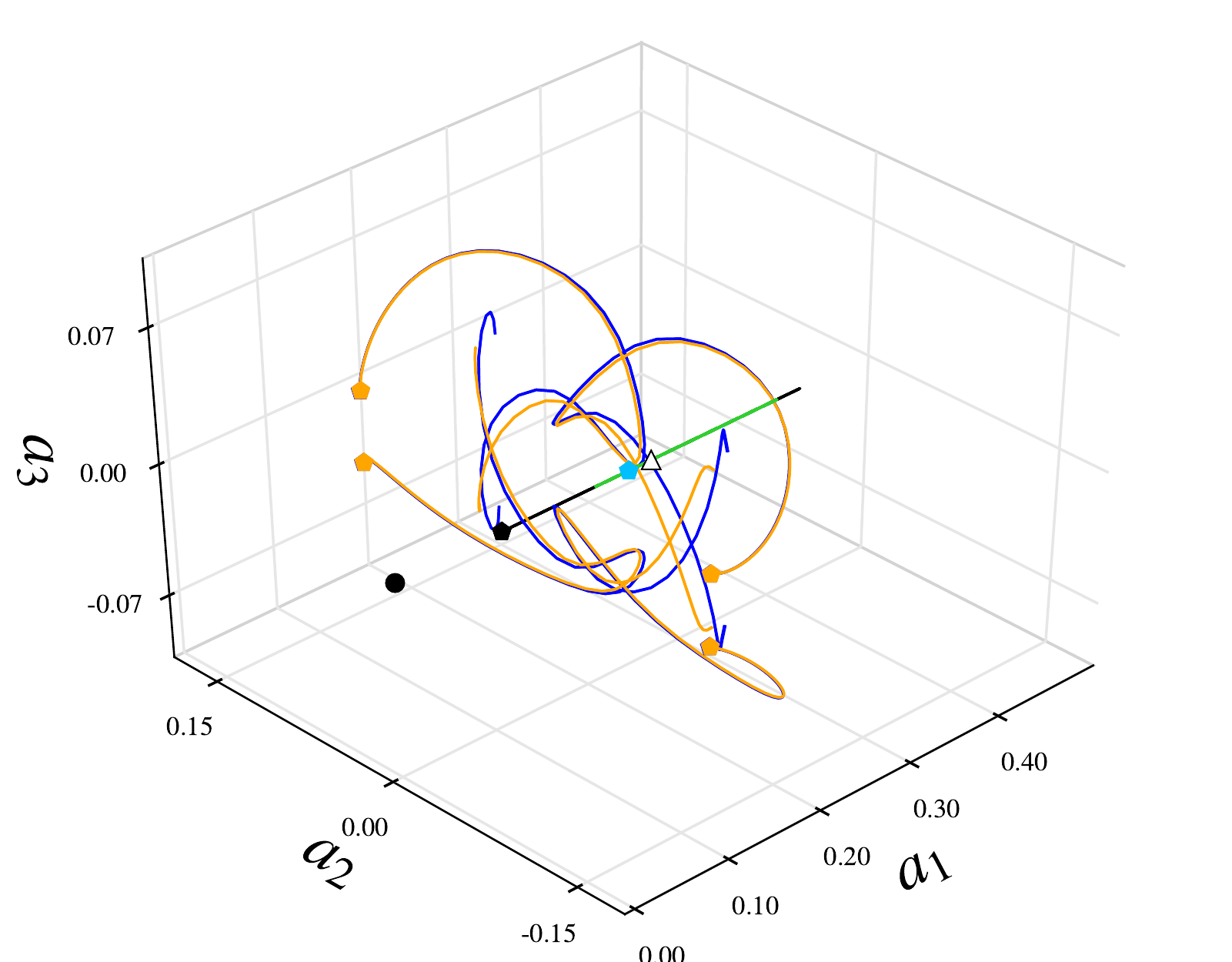}
			\caption{}
		\end{subfigure}
		\begin{subfigure}{0.42\linewidth}
			\centering
			\includegraphics[width=\linewidth]{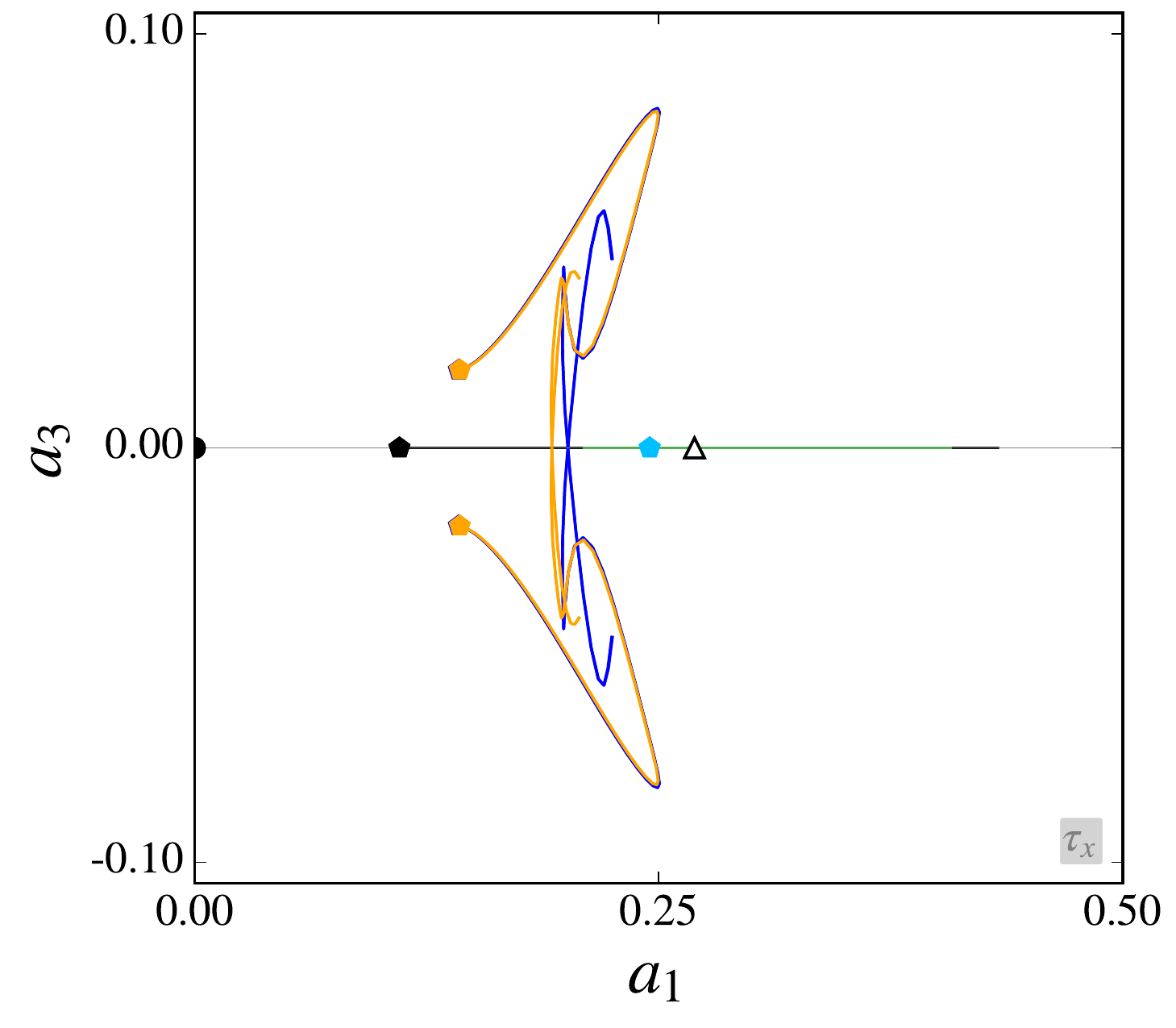}
			\caption{}
		\end{subfigure}
	\end{minipage}
	\begin{minipage}{.65\textwidth}
		\centering
		\begin{subfigure}{\linewidth}
			\centering
			\includegraphics[width=\linewidth]{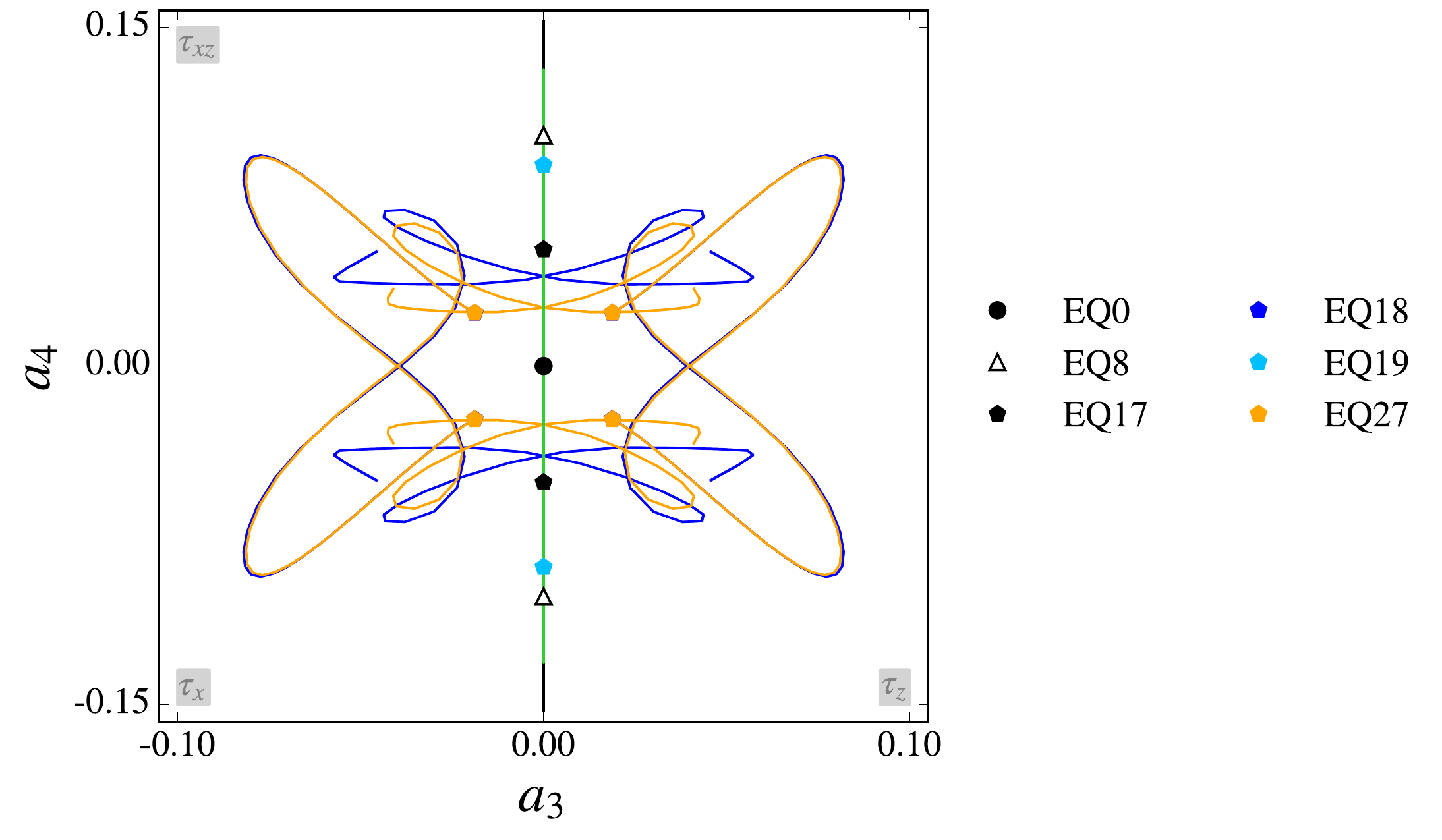}
			\caption{}
		\end{subfigure}
	\end{minipage}
	\caption{Equilibria discovered at $Re=270$ in state space. The previously known equilibrium solution (EQ8) is plotted with a triangular marker and the new equilibria are plotted with pentagonal markers. 
	The trajectories emanating from the markers represent fields that have been perturbed in the positive and negative eigendirections in state space.}
	\label{fig:re270_state_space}
\end{figure}

In order to understand the inter-relations between equilibria and to study their manifolds we must visualise the solutions in an appropriately reduced state space.
To this end, we use the representation independent state space projections described in \S\ref{sec:Statepace}.
Figures \ref{fig:re400_state_space}, \ref{fig:re330_state_space} and \ref{fig:re270_state_space} show the equilibria in state space at the distinct Reynolds numbers of $400$, $330$ and $270$, respectively.
The different regions of the figures have labels in the corners to denote where translated siblings of the equilibria are projected.
For example, if we shift EQ3 in the streamwise direction it is reflected in the $a_4$ axis (see figure \ref{fig:re400_state_space_a});
this behaviour is due to the symmetry properties of $\bs{e}_4$ as it is anti-symmetric with respect to streamwise shifts, see \eqnref{eqn:state_space}.
The distances between equilibria in the state space portraits below relate to the differences in the magnitude of fluctuations, i.e. the $\mathscr{L}^2$-norms of the fixed points.
Therefore, the further a fixed point is from the laminar solution (the origin) the greater its deviation from the laminar flow field.

% Computing the usntable manifolds
The trajectories seen leaving the equilibria in figures \ref{fig:re400_state_space}, \ref{fig:re330_state_space} and \ref{fig:re270_state_space} depict the unstable manifolds of the equilibria.
We use the method outlined by \cite{Gibson2008} to compute the unstable manifolds.
In short, we perturb an equilibrium solution in the direction of its leading unstable eigenvector and integrate the perturbed field forward in time to determine the unstable manifold.
Mathematically the perturbed field is defined as 
\begin{equation}\label{eqn:perturbed_field}
	\bs{u}(0) = \bs{u}_{EQ} \pm \epsilon\bs{v}_{EQ},
\end{equation}
where $\bs{u}_{EQ}$ is an equilibrium solution, $\epsilon$ is a scaling constant and $\bs{v}_{EQ}$ is the unstable eigenvector.
If the leading unstable eigenvalues of a solution are a complex conjugate pair, $\epsilon$ is varied in the range $10^{-2}\geq\epsilon\geq 10^{-4}$, whereas if the leading unstable eigenvalue is positive real only $\epsilon = 10^{-4}$.
Note that the unstable eigenvector selected in \eqnref{eqn:perturbed_field} belongs to the solution's symmetry-invariant subspace, hence all trajectories eventually decay onto the laminar solution.

% Re = 400
Focussing on figure \ref{fig:re400_state_space} first, we can see that EQ20, EQ21 and EQ22 have filled in the region between EQ1 and EQ10.
% EQ20/21/22
Those three new equilibria are close to the $a_1$ axis, which means that they do not vary as much when they are translated in the streamwise direction.
They do, however, vary significantly in the spanwise direction as is evident from figure \ref{fig:re400_state_space_e}.
The unstable manifolds of EQ20 and EQ21 form surfaces since the leading unstable eigenvalues are a complex conjugate pair (see figure \ref{fig:eigs}) and their perturbed trajectories spiral outwards from the fixed points. 
However, there are some trajectories of EQ20 that are directed towards the laminar solution --- those trajectories closely follow the manifold of EQ1 towards the laminar solution.
% EQ23/24
EQ23 and EQ24 sit near EQ2 and EQ11 respectively, and their unstable manifolds are not as locally complicated as expected when looking at the dimensionality of their unstable manifolds in table \ref{tab:equilibria}; 
the two new manifolds closely follow the shape of EQ11's manifold.
Both EQ23 and EQ24's unstable manifolds should form lines, however when perturbed in the negative eigendirection the trajectories swing back around to follow the dominant positive eigendirection.

% Re = 330
Figure \ref{fig:re330_state_space} shows that all solutions found from EQ6 are closer to the laminar solution than EQ6.
This matches the trends seen in table \ref{tab:equilibria}, where all child equilibria of EQ6 have a lower $\| \tilde{\bs{u}} \|$ and lower dissipation rate.
All solutions near EQ6 (including EQ6) are highly unstable with complicated manifold shapes that closely follow each other at certain points and generally resemble each other's manifold shape.
% EQ12/13/25
EQ12, EQ13 and EQ25 belong to the EQ7-8, EQ1-2 and EQ20-21-23 solution branches (see figure \ref{fig:bifs_all}).
Interestingly, the unstable manifolds of EQ13 and EQ25 are more unstable than the solutions that belong to their respective branches at $Re=400$, i.e. EQ1 and EQ20 in figure \ref{fig:re400_state_space}.
The unstable manifolds push trajectories away from the laminar solution at $Re=330$, which is in contrast to the behaviour seen at $Re=400$.
The manifold of EQ12 encases all of the other manifolds, but because of EQ12's symmetry subgroup, its manifold is only visible in the plots with $a_4$ as one of the axes. 
%This is because EQ12 is invariant under the $\tau_{xz}$ spatial shift.
The unstable manifolds of EQ6, EQ14, EQ15, EQ16 and EQ26 are quite complicated in shape and seem to be entangled with each other.
EQ6 and EQ15's manifolds initially follow each other very closely, eventually going their separate ways but maintaining a loose resemblence.

% Re = 270
Figure \ref{fig:re270_state_space} shows that EQ8's child equilibria are closer to the laminar solution.
% EQ8
EQ8's leading unstable eigenvalues are a complex conjugate pair indicating that a surface will be formed in state space.
However, the perturbed trajectories converge to almost one line that loops around and takes aim at the laminar solution.
Hence any turbulent trajectory near EQ8 would quite quickly be guided towards the laminar solution.
% EQ17 
EQ17's unstable manifold is similar to EQ8's, except it has a larger turning radius.
% EQ18/27
EQ18 is hidden behind EQ27, both solutions belong to a closely related pair of families, EQ18 to the EQ1-2 and EQ27 to the EQ20-21-23 branch, which originates from a pitchfork bifurcation from the EQ1-2 lower branch at approximately $Re=270$, see \S\ref{sec:eq_bifurcations}.
The unstable manifolds of EQ18 and EQ27 are more unstable than other solutions in their respective families at higher Reynolds numbers since the perturbed trajectories stray quite far from their respective fixed points.
The manifolds of EQ18 and EQ27 follow each other closely until they diverge after crossing into opposing quadrants.
As noted in table \ref{tab:equilibria}, EQ19 does not have any unstable eigenvalues in the symmetry-invariant subspace, thus no manifold could be computed.


\subsection{Continuation under $Re$}\label{sec:eq_bifurcations}
% Naming convention for solutions on one bifurcation curve (solution-grouping)
The solutions were parametrically continued with $Re$ as the bifurcation parameter and subsequently grouped into pairs by observing their bifurcation curves and identifying which part of the curve each solution inhabits.
The naming convention for branch pairs is EQX-Y, where X indicates the `lower' branch solution (lower dissipation rate) and Y signifies the `upper' branch solution (higher dissipation rate).
There is one case where three equilibria sit on one curve, in that case the naming convention is in ascending order of dissipation rate of each equilibrium solution.

\begin{figure} % BRANCHES All
	\centering
	\includegraphics[scale=0.5]{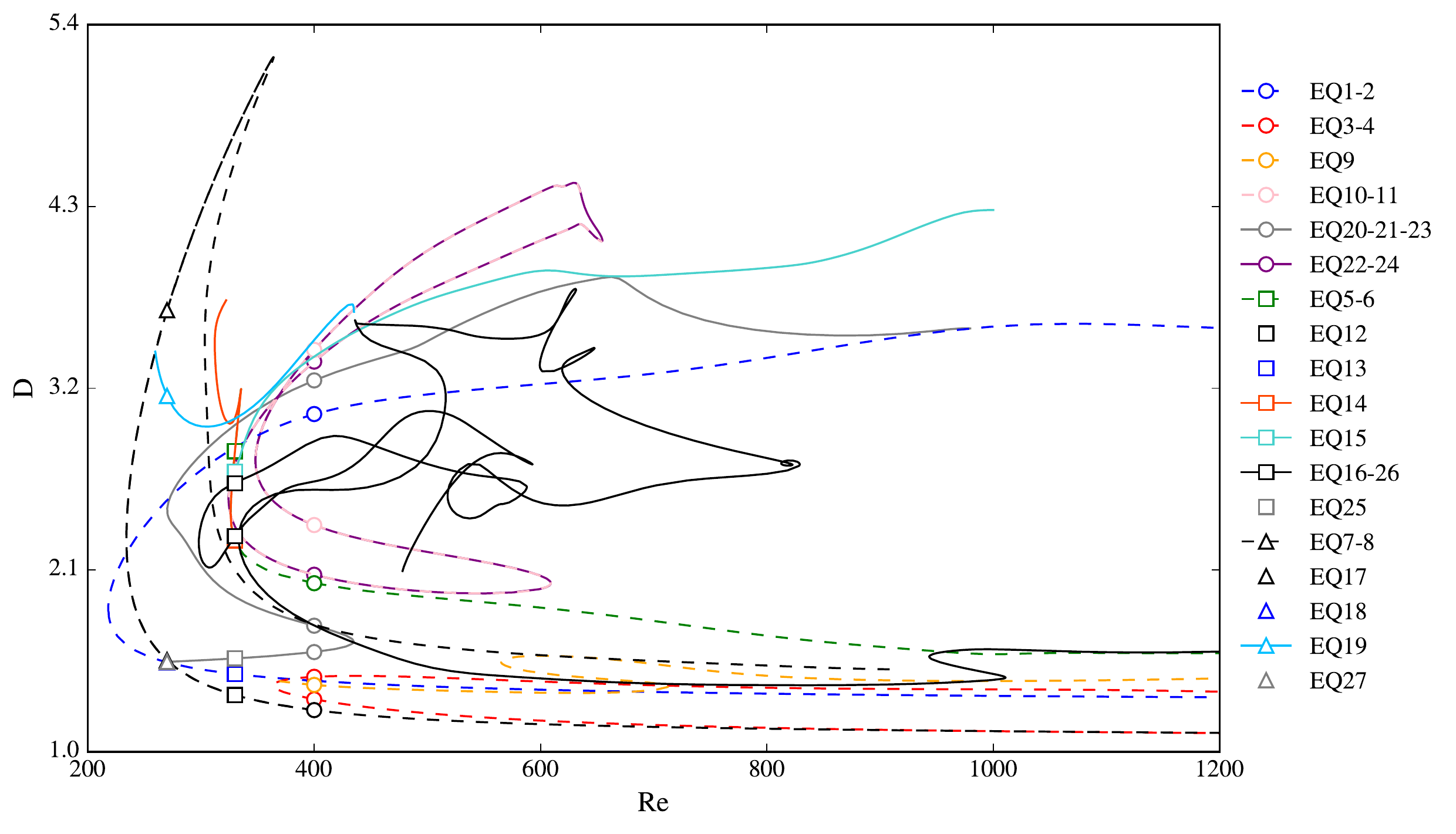}
	\caption{Dissipation of equilibria as a function of $Re$.
		All equilibrium solutions continued upwards and downwards in bifurcation parameter $Re$, starting from the $Re$ they were discovered at signified using the following symbols: $\Circle$ for $Re=400$, $\square$ for $Re=330$ and $\triangle$ for $Re=270$.
		The marker colour matches the colour of the curve it belongs to.
		The dashed lines signify previously known branches of solutions, the solid lines are bifurcation curves found from continuing the new equilibria.}
	\label{fig:bifs_all}
\end{figure}

\begin{figure} % BRANCHES Re = 400
	\centering
	\includegraphics[scale=0.5]{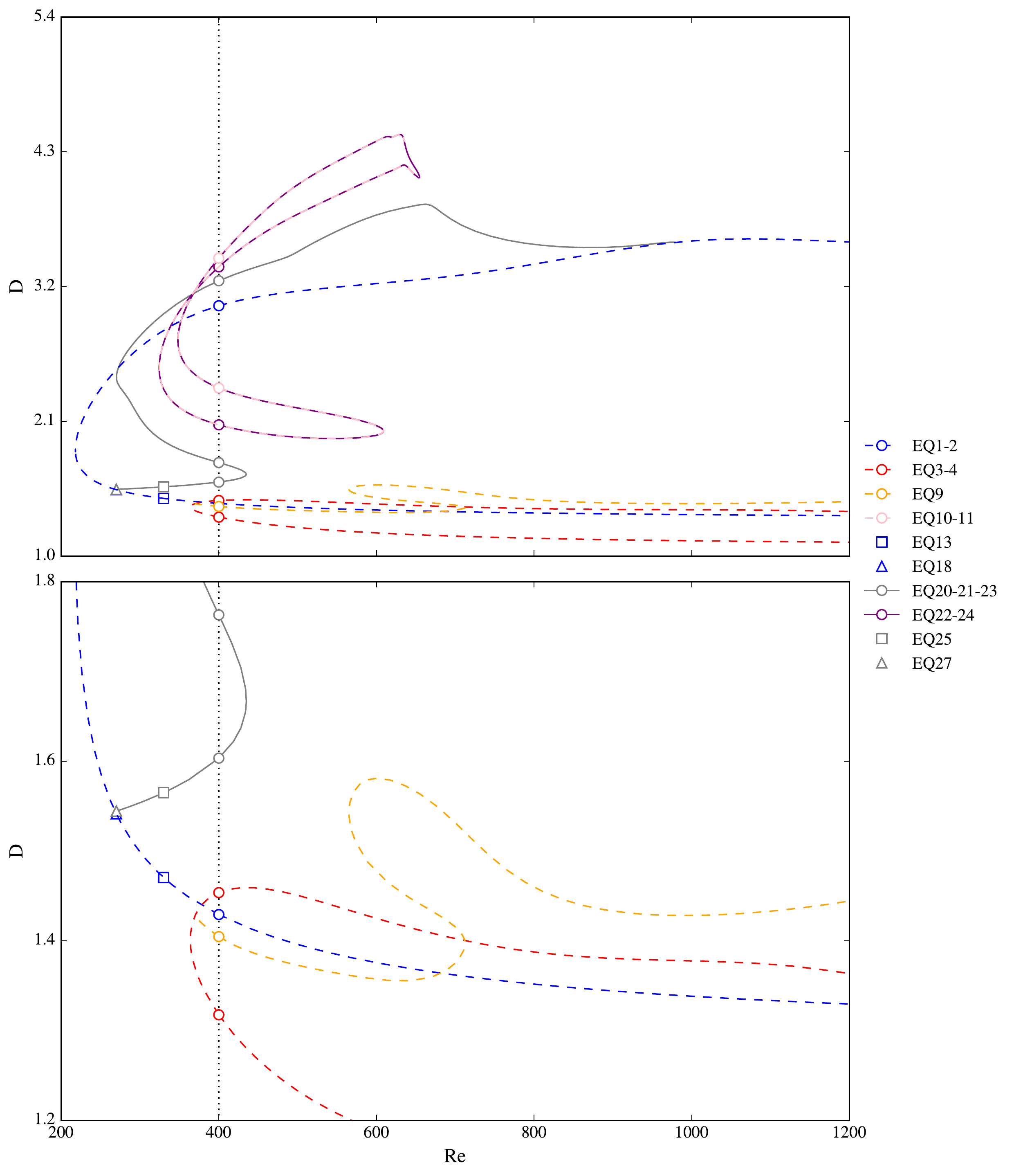}
	\caption{Bifurcation curves of equilibrium solutions discovered at $Re = 400$ continued upwards and downwards in bifurcation parameter $Re$.
	The second figure is a zoomed-in plot of the pitchfork bifurcations.}
	\label{fig:bifs_400}
\end{figure}

\begin{figure} % BRANCHES Re = 330
	\centering
	\includegraphics[scale=0.5]{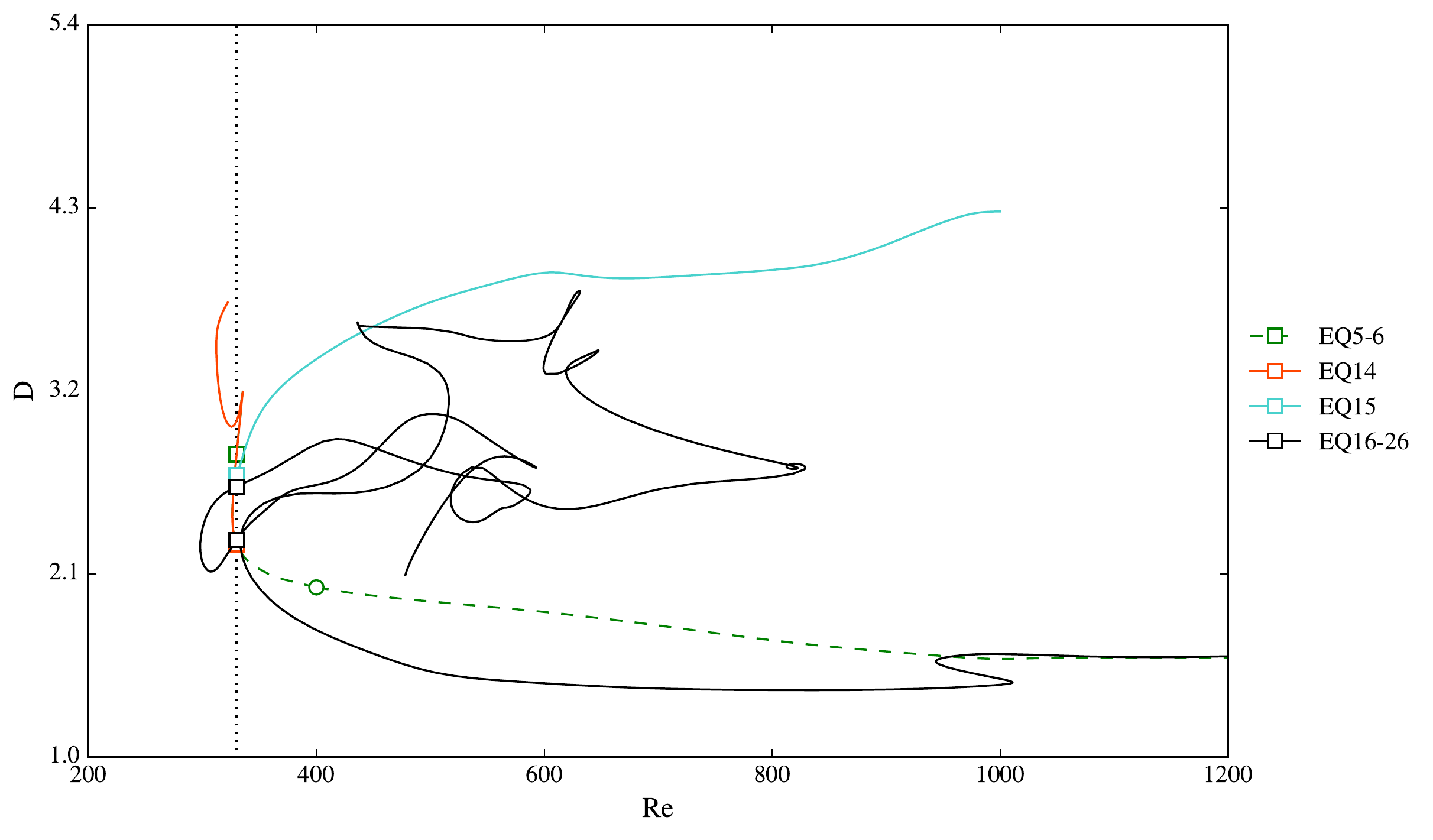}
	\caption{Bifurcation curves of equilibrium solutions discovered at $Re = 330$ continued upwards and downwards in bifurcation parameter $Re$.}
	\label{fig:bifs_330}
\end{figure}

\begin{figure} % BRANCHES Re = 270
	\centering
	\includegraphics[scale=0.5]{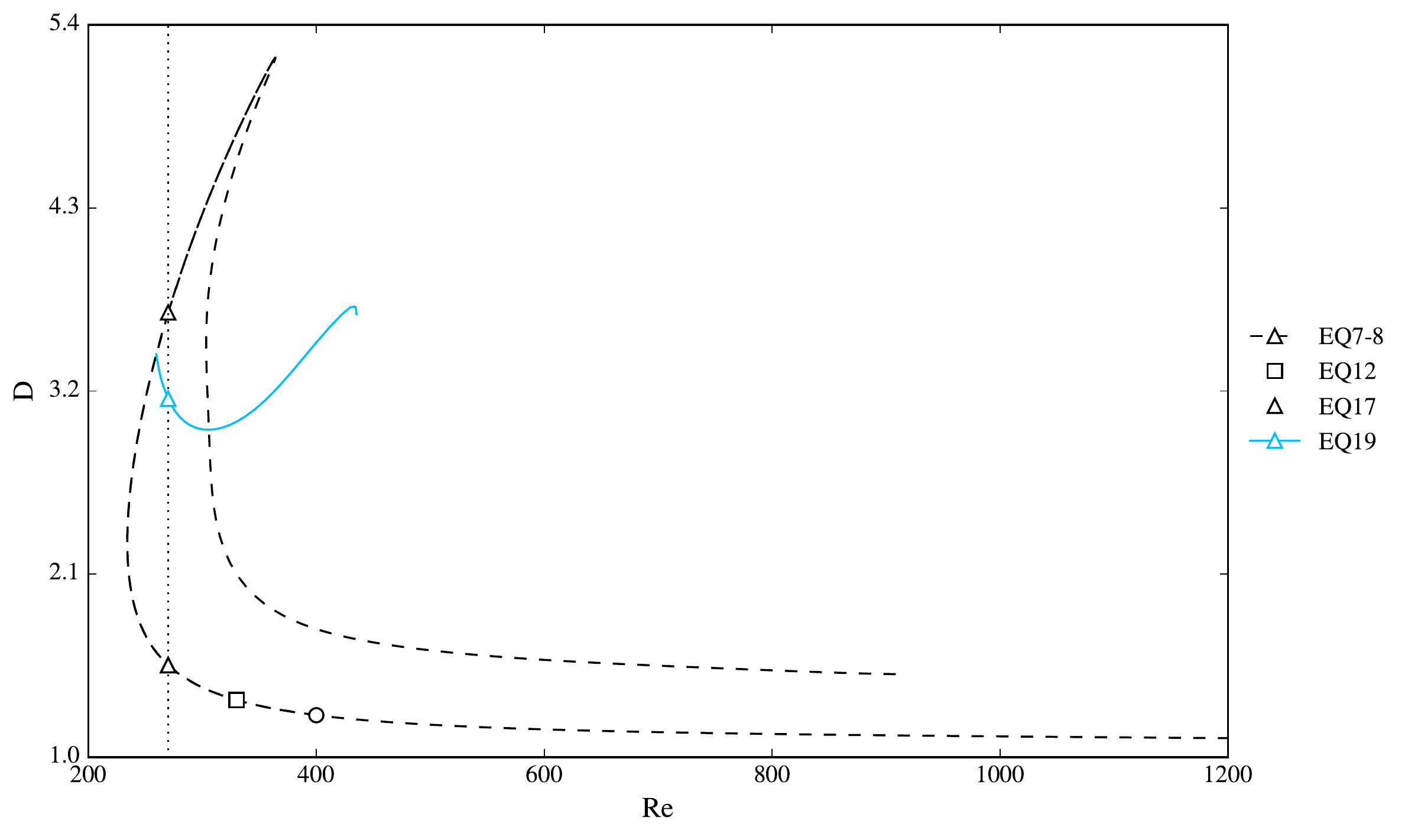}
	\caption{Bifurcation curves of equilibrium solutions discovered at $Re = 270$ continued upwards and downwards in bifurcation parameter $Re$.}
	\label{fig:bifs_270}
\end{figure}

% How is it plotted? What are you showing in your plots?
The bifurcation diagrams in figures \ref{fig:bifs_all}, \ref{fig:bifs_400}, \ref{fig:bifs_330} and \ref{fig:bifs_270} are plotted with the dissipation rate $D$ against the bifurcation parameter $Re$.
The curves are independent of each other and any apparent intersections between curves are not bifurcations, 
except in the case of the pitchfork bifurcations which are highlighted in the description below.
Each curve represents a family of solutions with an upper and lower branch, originating from bifurcations at certain Reynolds numbers.


\begin{figure} % 3D figures 1
	\centering
	\begin{subfigure}{0.8\linewidth}
		\centering
		\includegraphics[width=\linewidth]{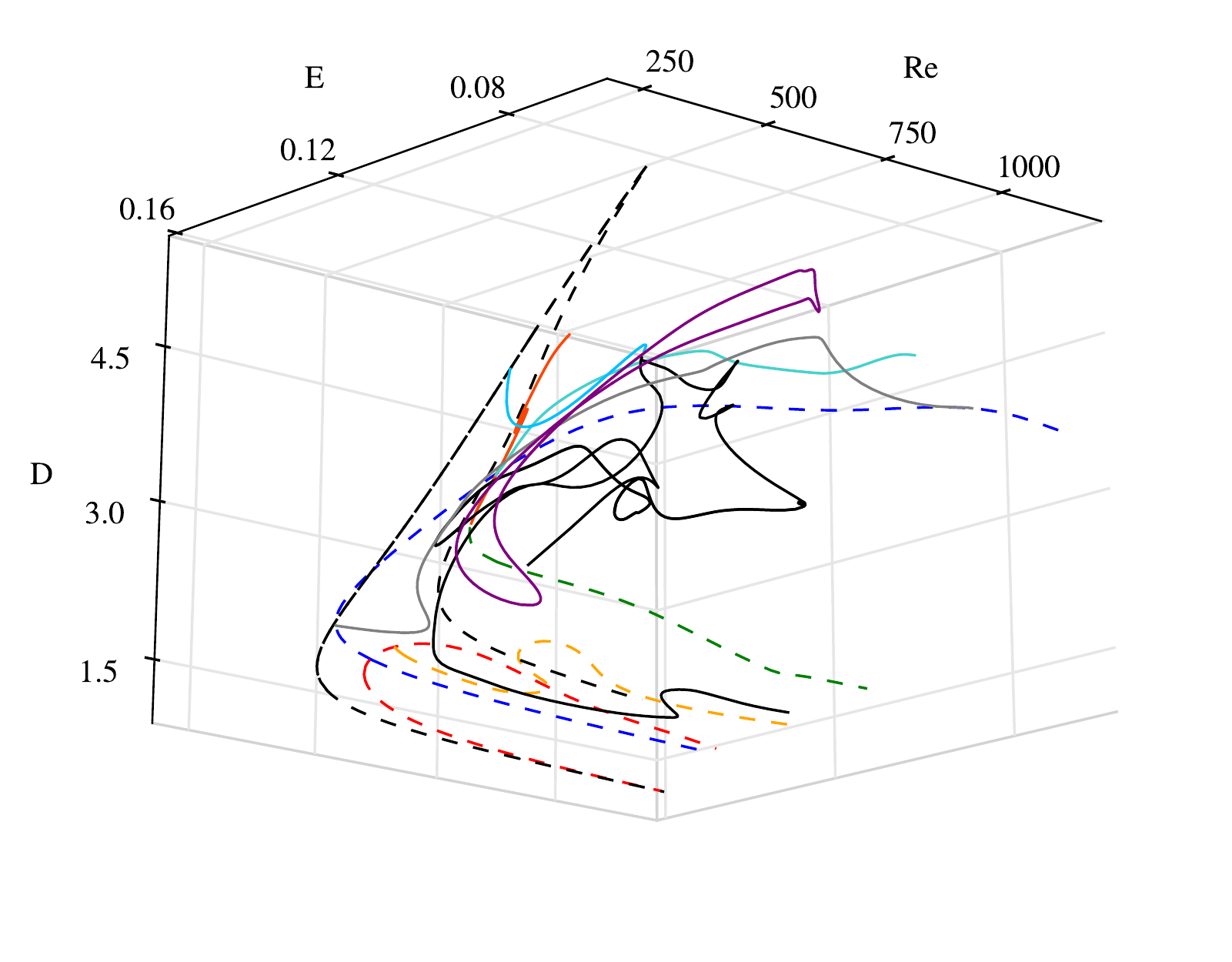}
		%		\captionof{figure}{}
	\end{subfigure}\\[-8ex]
	\begin{subfigure}{0.8\linewidth}
		\centering
		\includegraphics[width=\linewidth]{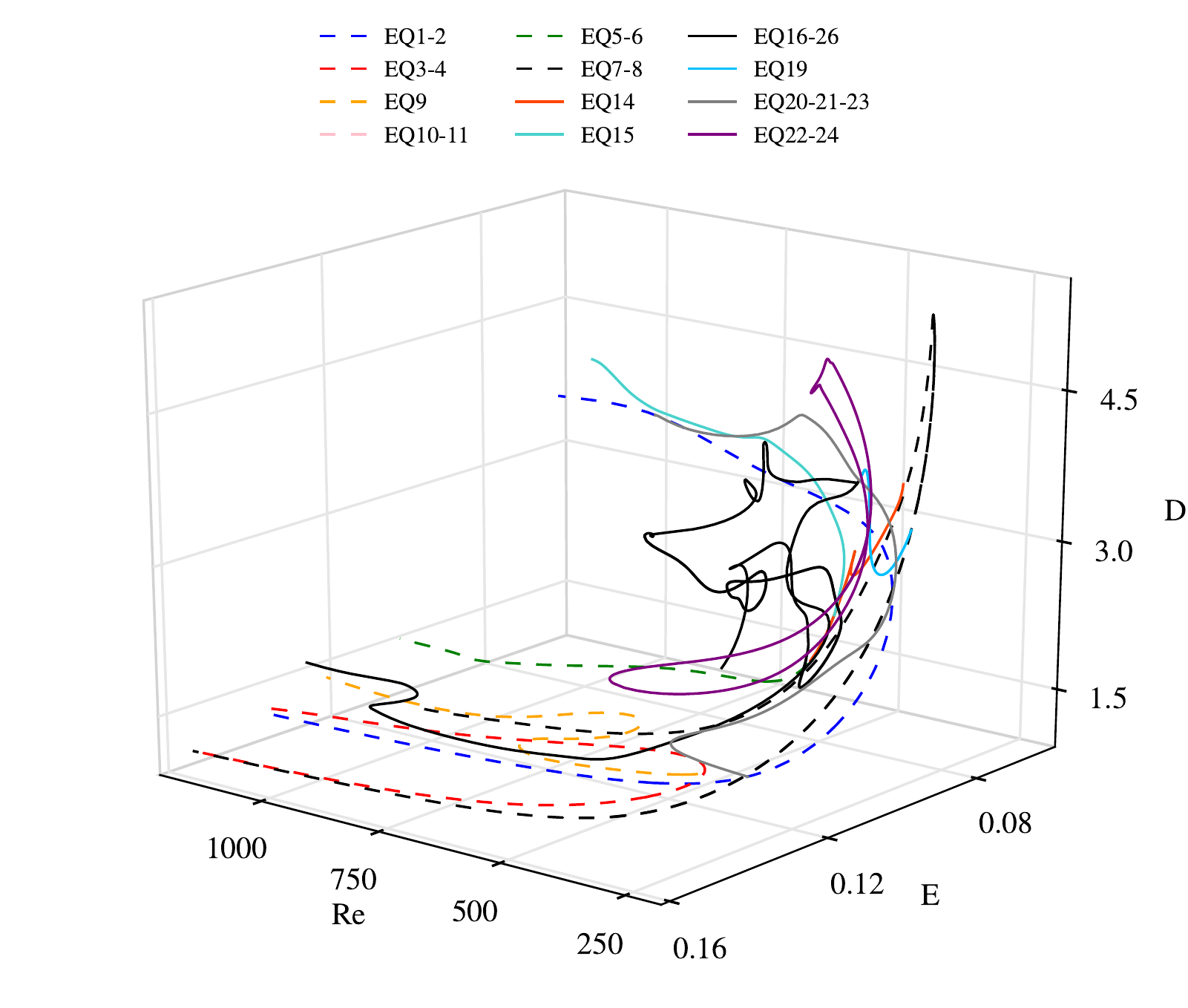}
		%		\captionof{figure}{}
	\end{subfigure}
	\caption{Three-dimensional plots looking at the relationship between the dissipation rate $D$, total energy density $E$ and Reynolds number $Re$. The dashed lines represent the bifurcation curves of previously known branches and the solid lines signify the new branches.}
	\label{fig:3D_bifs}
\end{figure}

Figure \ref{fig:3D_bifs} shows how the bifurcation curves vary with respect to kinetic energy density $E$ as well as $D$ and $Re$.
The figure shows that solutions with high $D$ have low $E$ and the bifurcations seem to roughly form a pseudo-hyperplane in their $Re-D-E$ projection.
We have already seen in the previous section that equilibria with large dissipation rates are typically further from the laminar solution. Since turbulence is highly dissipative, the less dissipative equilibria would seem to be more relevant in understanding transitional trajectories.

\subsubsection{Bifurcation curves of equilibria found at $Re = 400$}
% EQ1-EQ2
In ascending $Re$, the first saddle-node bifurcation encountered is at $Re=218.1$ which produces the \cite{Nagata1990} EQ1-2 branches, respectively.
% EQ20-21-23
The EQ20 branch originates from a pitchfork bifurcation at $Re=268.5$ from \citeauthor{Nagata1990}'s lower branch, see figure \ref{fig:bifs_400}. 
There is a saddle-node bifurcation at $Re=270.2$ which forms the EQ21-23 curve that reverse bifurcates with the EQ20 branch at $Re=434.9$, forming the EQ20-21-23 curve, which merges with the EQ2 branch at $Re=980.4$.

% EQ22-24
The EQ10-11 branches merge with another bifurcation curve at $Re=609$ and almost form a loop, the curves remain disconnected between $Re\approx635-655$, as shown in figure \ref{fig:bifs_400}.
When EQ22 and EQ24 are continued in $Re$, it is found that they inhabit the lower and upper branches of the curve that EQ10-11 merges with, respectively.
It is also found that when continued, EQ22 and EQ24 close the aforementioned incomplete loop.
There are no symmetry changes observed for any solution in the bifurcation loop and the flow field structure is similar across both sets of branches.
The EQ10-11 branches come from a saddle-node bifurcation at $Re=348.2$,
and the EQ22-24 branches originate from a saddle-node bifurcation at $Re=324.3$.

% EQ9
The EQ9 branch originates from a pitchfork bifurcation at $Re=369.3$ from the EQ3-4 curve as shown in figure \ref{fig:bifs_400}.
When continued up in Reynolds number, the EQ9 branch undergoes a reverse saddle-node bifurcation at $Re=712.8$ ($D=1.4$) followed by a saddle-node bifurcation at $Re=565.1$ ($D=1.54$), after which it smoothly continues to higher Reynolds numbers.

\subsubsection{Bifurcation curves of equilibria found at $Re = 330$}
Some of the new solutions are situated on branches that were previously known, 
hence when the child equilibria are continued in Reynolds number it leads to the rediscovery of a branch that was already known. 
This can be seen in figures \ref{fig:bifs_400}, \ref{fig:bifs_330} and \ref{fig:bifs_270} when symbols signifying different Reynolds numbers appear on the same curve; 
different symbols on the same curve are indicative of a branch-jump.

% EQ14
When EQ14 is continued down in Reynolds number, the bifurcation curve follows the path to EQ6 and continues further to form another branch.
The curve increases to a dissipation rate of 3.2 at $Re=335.4$ and then rapidly but smoothly decreases in dissipation rate with decreasing Reynolds number towards a saddle-node bifurcation at $Re=312.2$.
The two bifurcation curves merge at $Re=335.4$, note that there is no bifurcation here as there are no symmetry or stability changes at this point.

% EQ15
The EQ5-6 curve also has a pitchfork bifurcation that takes place at $Re = 328$, leading to EQ15.
The bifurcation is accompanied by a symmetry change from $\Sigma$ to $\Omega_3$ for all solutions that inhabit the new branch.

% EQ16, EQ26
Lastly, the EQ16-26 bifurcation curve starts from a saddle-node bifurcation at $Re=298.1$.
It consists of a lower-upper branch pair of solutions that cross in dissipation at $Re=453.7$, so that the lower and upper branches become the upper and lower branches, respectively. 
The upper branch then smoothly curves back on itself; the dissipation rate decreases followed by a slight increase as it loops back on itself while $Re$ increases.
After that both $D$ and $Re$ decrease almost linearly, until the curve stops at $Re=478.1$, $D=2.1$.
Focussing on the lower branch, the bifurcation curve then merges with another bifurcation curve at $Re=587.9$, which originates from a saddle-node at $Re=517.9$.
The merged curve then continues upwards in Reynolds number until $Re=822.7$, at this point the curve loops and continues downwards in Reynolds number. 
There are three further loops where the symmetry of the solutions change from $\Theta$ to $K$ indicating that a bifurcation has taken place.
The final part of the curve goes through a saddle-node bifurcation at $Re=436.1$, $D=3.61$ and continues upwards in $Re$ whilst maintaining a very low dissipation rate.
There is a further merging with another bifurcation curve at $Re=1010.9$ that is spawned from a saddle-node bifurcation at $Re=943.3$.
This curve does not exhibit any twists and turns beyond $Re=1200$ and note that in figure \ref{fig:bifs_330} it seems as though this curve joins with the EQ5-6 lower branch, however this is not the case as can be seen in figure \ref{fig:3D_bifs}.
It should be noted that at $Re=478, D=2.1$ and $Re=450, D=3.4$ there are unstable fixed points that, when approached, cause the continuation algorithm to reverse direction and retrace the path it took to those points.

\subsubsection{Bifurcation curves of equilibria found at $Re = 270$}
% EQ7-EQ8 curve
The EQ7-8 branches are formed from a saddle-node bifurcation at $Re=234.2$, see figure \ref{fig:bifs_270}.
At $D=5.2$ the curve merges with another bifurcation curve; this is evident from a rapid and smooth continuation at $Re=364.2$.
A lower-upper branch pair specifically at $Re=270$ can be formed with EQ17-8, 
and EQ7 can be paired with the corresponding solution at $Re=400$ on the upper branch.

% EQ19 curve
The EQ19 curve originates from a pitchfork bifurcation at $Re=259.6$. % Bifurcation between search-72-73
The symmetry of the equilibria change at the bifurcation point, going from EQ19 to EQ7 the symmetry subgroup changes from $K$ to $\Theta$. 
The curve reaches a maximum dissipation at $D=3.71$ at $Re=433$, then experiences a slight decrease in dissipation to $D=3.66$ at $Re=435.2$.

% EQ18 and EQ27
As mentioned before, EQ18 and EQ27 are points on the bifurcation curves for the EQ1-2 and EQ20-21-23 branches, respectively.
Hence, when the new equilibria are continued in Reynolds number their curves are identical to the aforementioned parent branches.
EQ12 (discovered at $Re=330$) is the result of a branch-jump from the EQ5-6 curve to the EQ7-8 curve, see figure \ref{fig:bifs_270}.

\section{Discussion}\label{sec:Discussion}
The results in \S\ref{sec:Results} reveal that, in most cases, equilibria with less fluctuation energy and lower dissipation (compared to their parent solutions) can be found using the project-then-search method.
We speculate that the project-then-search technique allows us to take known equilibria on higher-dimensional unstable manifolds and use their projections (which are close to lower-dimensional manifolds) to find equilibria that sit on lower-dimensional unstable manifolds.
The low-dimensional nature of the new equilibria is attributed to the projection generated by the resolvent model; resolvent modes for a given equilibrium solution span a low-dimensional space which the solution approximately inhabits.
This is evident when comparing the dimensionality of the unstable manifolds of child equilibria against their parent solutions;
almost all second/third generation solutions have an unstable manifold with reduced dimensionality in both the symmetry-invariant subspace and the full space.
Also, it is clear from table \ref{tab:equilibria} that there is a relationship between a solutions' dissipation rate, energy and the dimensionality of its unstable manifold in both the symmetry-invariant subspace and the full space. 
In general, the higher the dissipation rate and fluctuation magnitude are, the more unstable the solution's manifold is.

\begin{figure} % Successful Searches
	\centering
	\includegraphics[width=\textwidth]{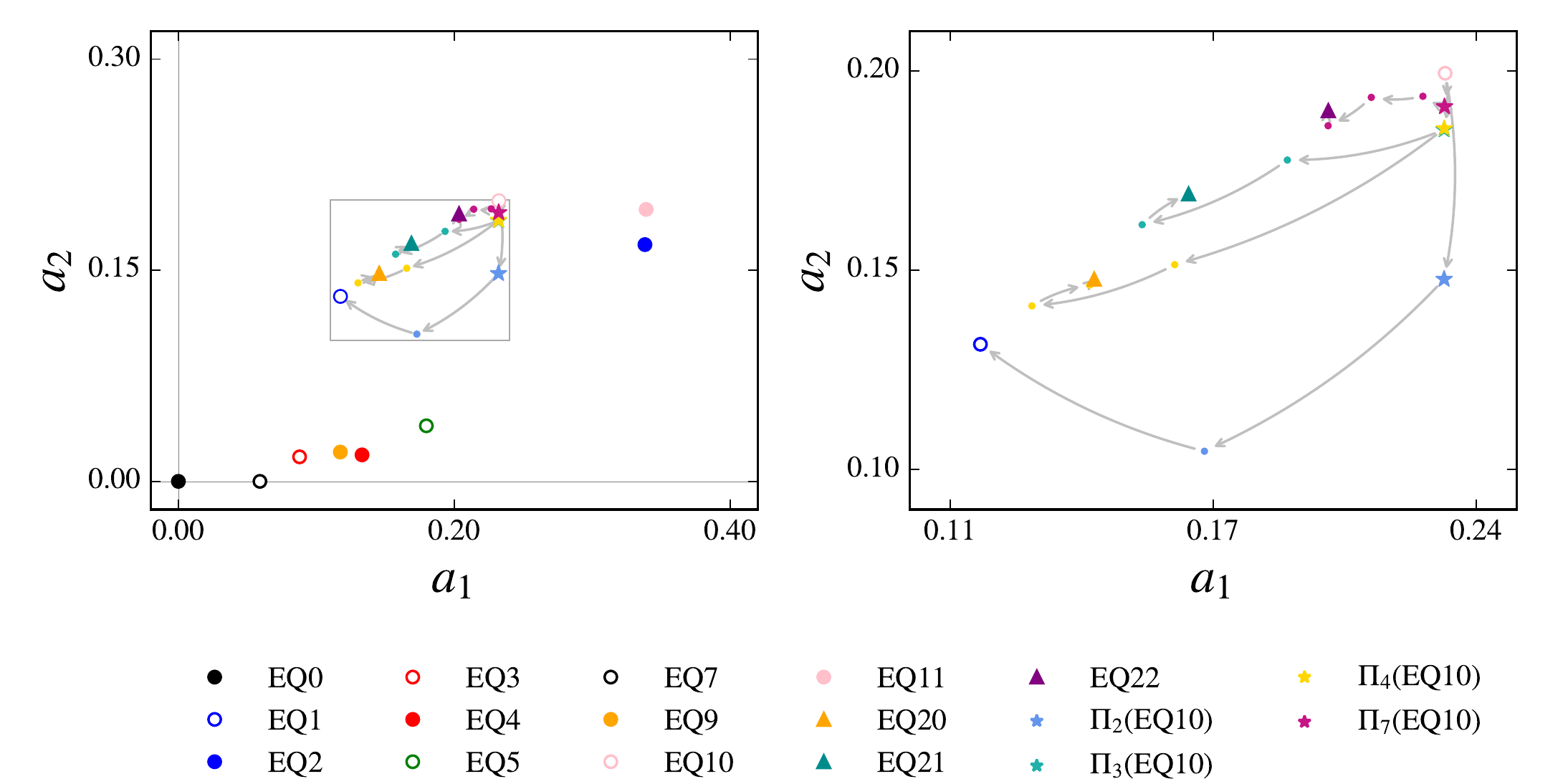}
	\caption{A state space representation of successful searches initiated from low-rank projections of EQ10. 
		The plot on the right shows the detail of the grey box in the left figure.}
	\label{fig:search_successful}
\end{figure}

To illustrate the mechanics of the project-then-search method, the process can be shown in state space.
For the purposes of demonstration, successful searches from EQ10's low-rank projections are shown in figure \ref{fig:search_successful}.
The search algorithm is quick to locate the region where a potential solution exists.
Once the search is very close to a solution, it takes just a few Newton steps (not shown) to converge onto the solution.
It is clear from figure \ref{fig:search_successful} that small differences in the initial state used for the NKH algorithm can lead to very different results, indicating that the initial conditions are close to a basin boundary.
For example $\Pi_3$(EQ10) and $\Pi_4$(EQ10) are extremely close to each other in state space, yet they lead to two different equilibria.

\begin{figure} % EQ10 flow ranked-approximations images
	\centering
	\begin{minipage}{\textwidth}
		\centering
		\begin{subfigure}{\linewidth}
			\centering
			\includegraphics[width=.4\linewidth]{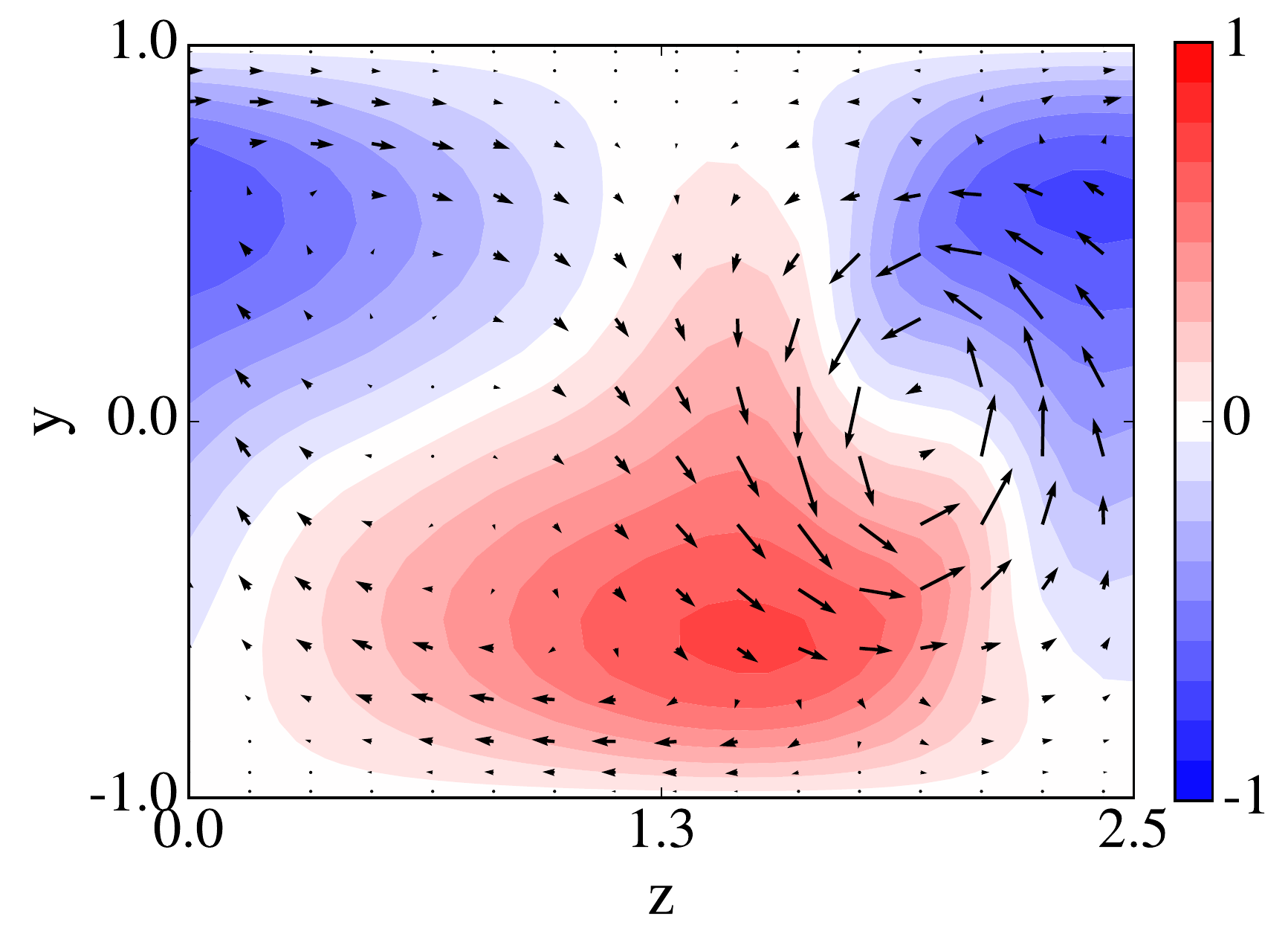}
			\caption{EQ10}
		\end{subfigure}
	\end{minipage}\\
	\begin{minipage}{.38\textwidth}
		\centering
		\begin{subfigure}{\linewidth}
			\centering
			\includegraphics[width=\linewidth]{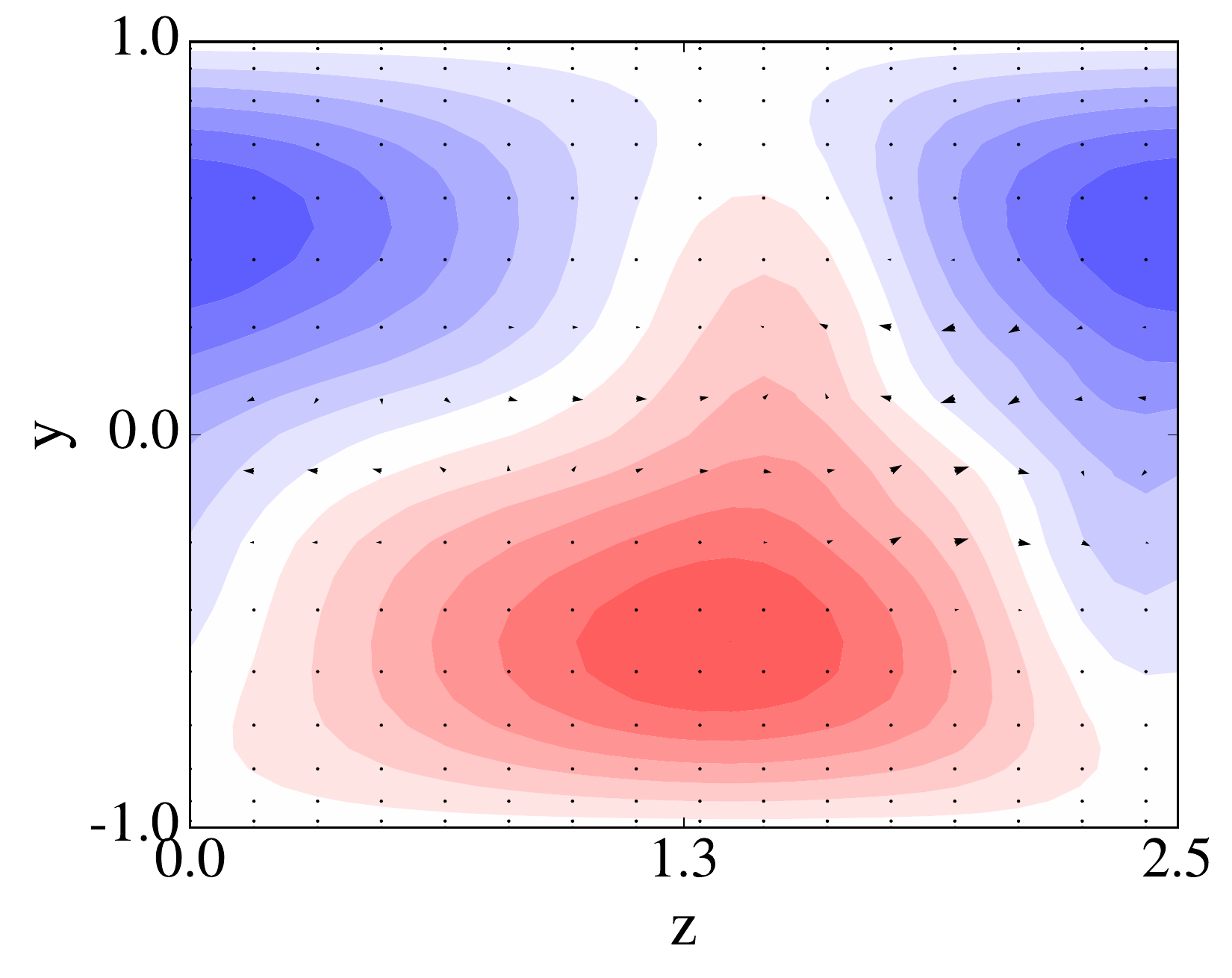}
			\caption{\rank{3}{10}}
		\end{subfigure}\\[1ex]
		\begin{subfigure}{\linewidth}
			\centering
			\includegraphics[width=\linewidth]{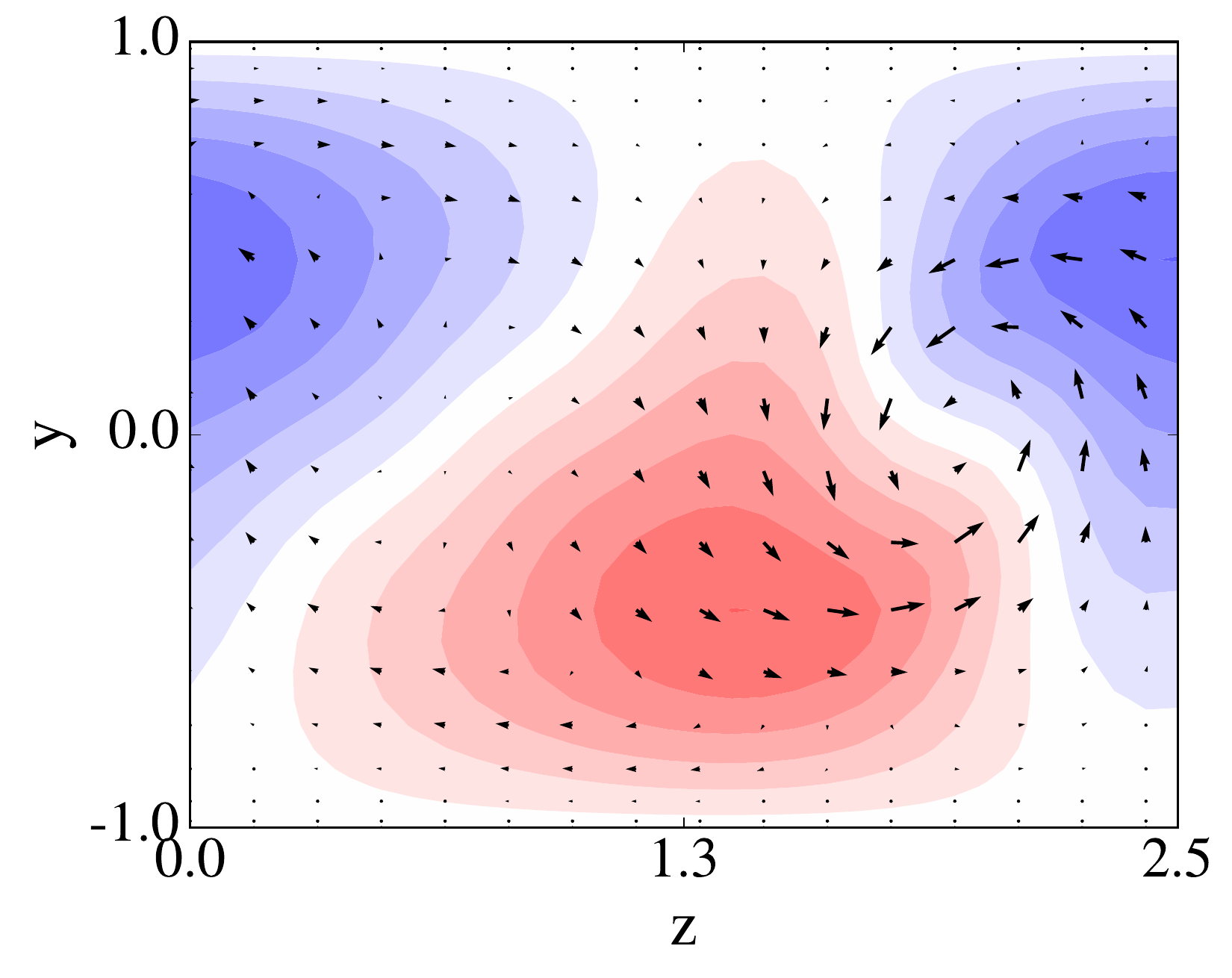}
			\caption{EQ21}
		\end{subfigure}
	\end{minipage}%
	\begin{minipage}{.38\textwidth}
		\centering
		\begin{subfigure}{\linewidth}
			\centering
			\includegraphics[width=\linewidth]{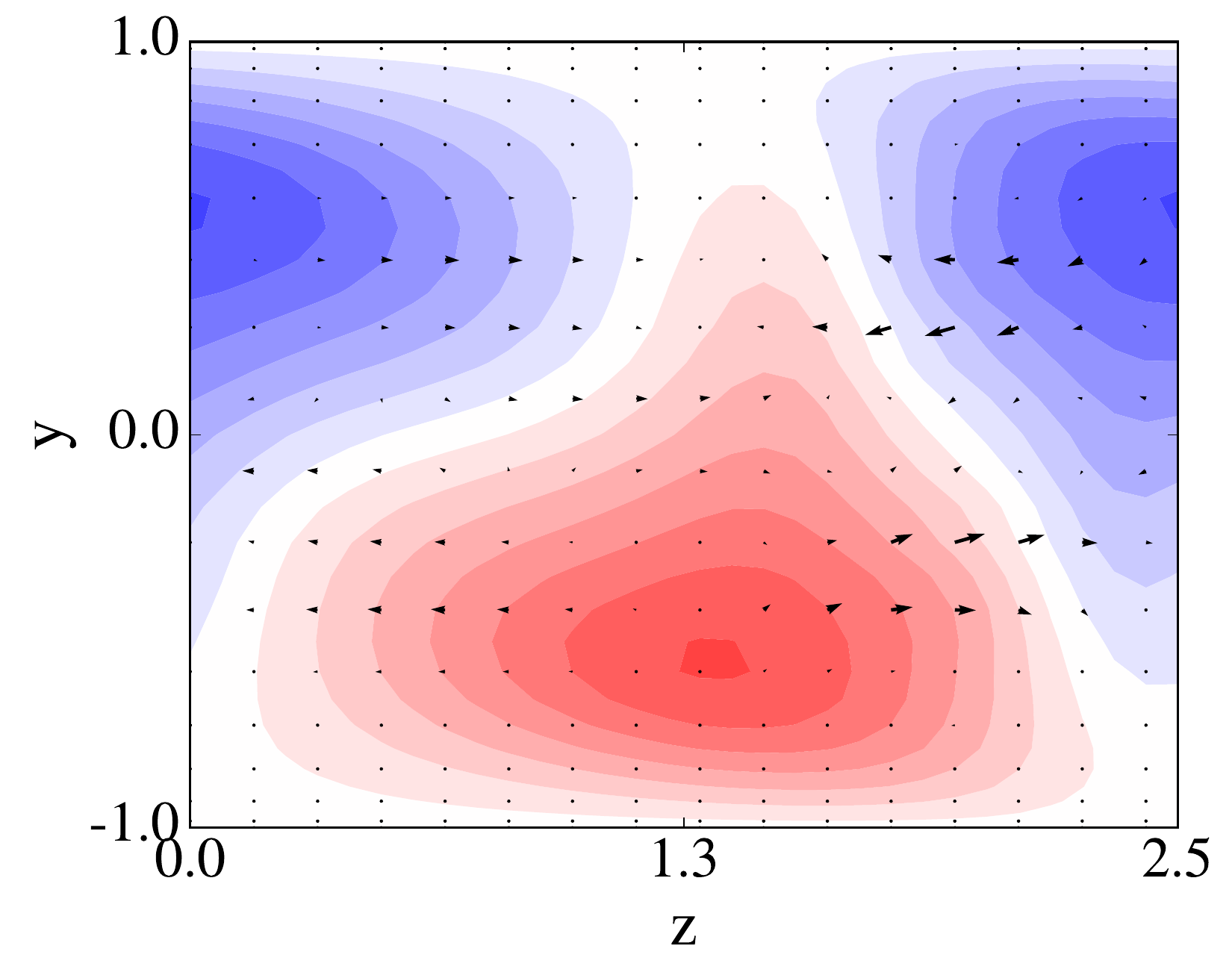}
			\caption{\rank{7}{1}}
		\end{subfigure}\\[1ex]
		\begin{subfigure}{\linewidth}
			\centering
			\includegraphics[width=\linewidth]{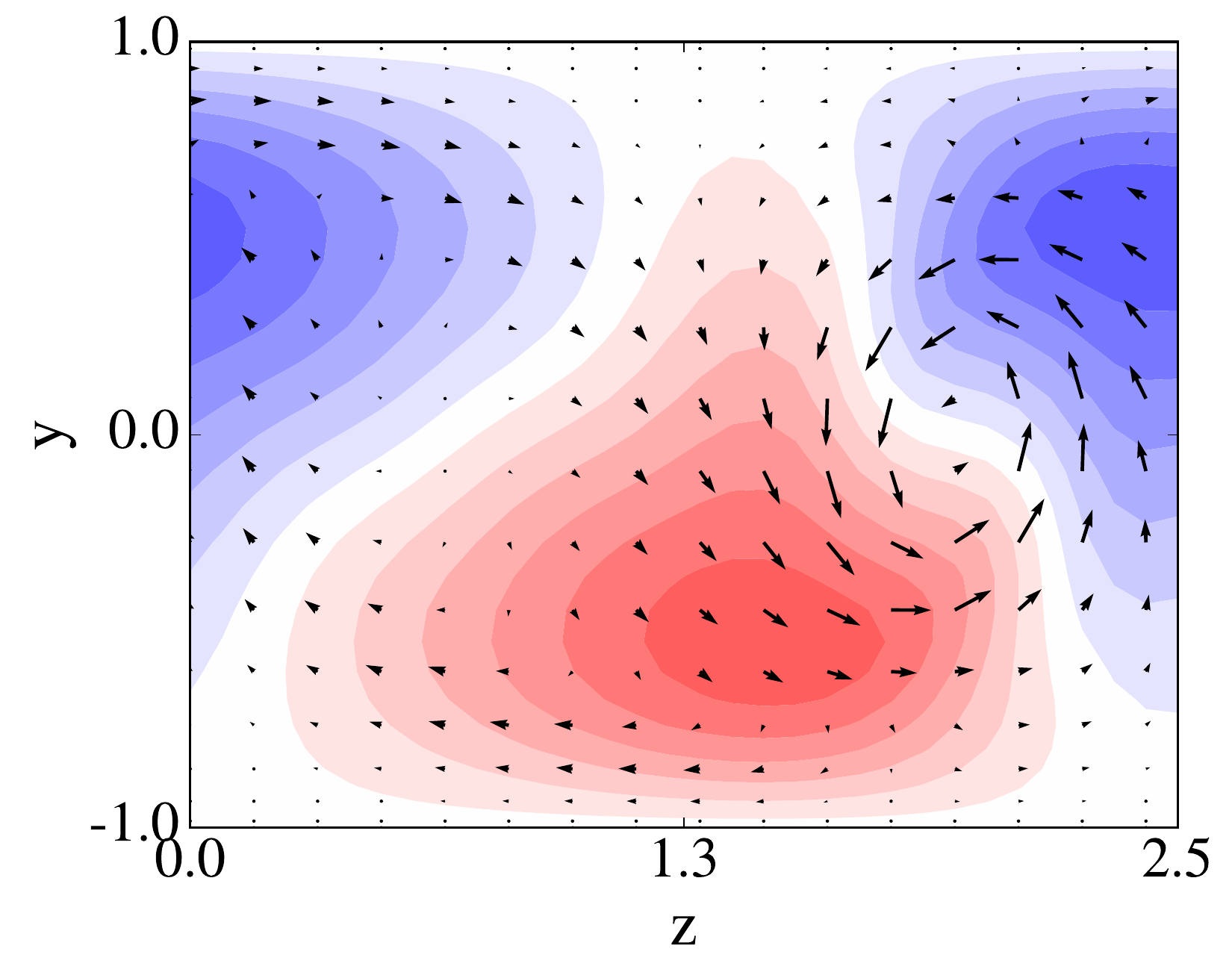}
			\caption{EQ22}
		\end{subfigure}
	\end{minipage}
	\caption{Plots of (a) EQ10 and two of its projections (b) and (d) that lead to (c) EQ21 and (e) EQ22, all plots are at $x=0.0$. 
		The fields shown are differences from laminar, 
		i.e. $\tilde{\bs{u}} = \bs{u} - y\hat{\bs{x}}$, 
		where the arrows show $[\tilde{v}, \tilde{w}]$ and the arrow lengths represent the velocity magnitude at a particular point in-plane. 
		The colours red/blue indicate $\tilde{\bs{u}}=\pm1$, and white indicates $\tilde{\bs{u}}=0$.
}
	\label{fig:flow_fields}
\end{figure}

Visualising the project-then-search process in the reduced state space allows us to see the distance and direction of the projections relative to the original solution.
Figure \ref{fig:search_successful} shows that lowering the rank produces projections that are simultaneously further form the original solution and closer to the laminar state, in an energy sense.
This is to be expected as the resolvent modes used to generate the projections are derived with respect to an energy norm.
As the resolvent modes are truncated during the projection process, the small length-scale content of the original field is removed and energetic large-scale motions remain.
The resulting fields of varying rank are `smoothed' versions of the original solution,
thus, the projected fields tend to be less dissipative.
The projected fields physically resemble the original fields but with lower fluctuation magnitude (see figure \ref{fig:flow_fields}).
Consequently, less dissipative equilibria are discovered, with EQ20 and EQ27 being exceptions; 
they have a higher dissipation rate and energy magnitude than their parent.
By way of example, the flow fields of the projections and final solutions from figure \ref{fig:search_successful} are shown in figure \ref{fig:flow_fields}.
The plots clearly show that the projections remove small length-scales near the wall while maintaining the larger structures in the central region of the domain.

% All ranks were used
For every solution all available ranks were used to generate projections.
The `Root' column in table \ref{tab:equilibria} denotes the lowest ranked-projection that leads to the discovery of a solution, but this does not mean that that is the \emph{only} projection that yields that solution.
All ranked-projections of the equilibria lead to the discovery of other non-trivial unstable solutions, although there are some equilibria whose projections only converge to the laminar state or return to the original fixed point (this fact is denoted in table \ref{tab:equilibria} under the `Ret.' column).
Examples of such equilibria include EQ1, EQ3, EQ4, EQ7 and EQ20.
% Failed searches
There are, however, some projections that fail to initialise any successful search at all;
there were a total of 107 failed searches, which translates to a $4\%$ failure rate. 
For these cases the search was repeated with eight times as many Newton steps ($160$ rather than $20$), yet the solutions failed to converge;
\cite{Viswanath2007} found that for a system with $10^5-10^6$ unknowns, the NKH algorithm takes less than $100$ steps to find an exact solution.
In relation to \eqnref{eqn:NKH}, the required tolerance for successful searches is $ \| G\|  \sim 10^{-15}$, 
in contrast we find that for unsuccessful searches $10^{-6} \lesssim \| G\|  \lesssim 10^{-3}$.
\begin{figure} % UNSUCCESSFUL Searches
	\centering
	\includegraphics[width=\textwidth]{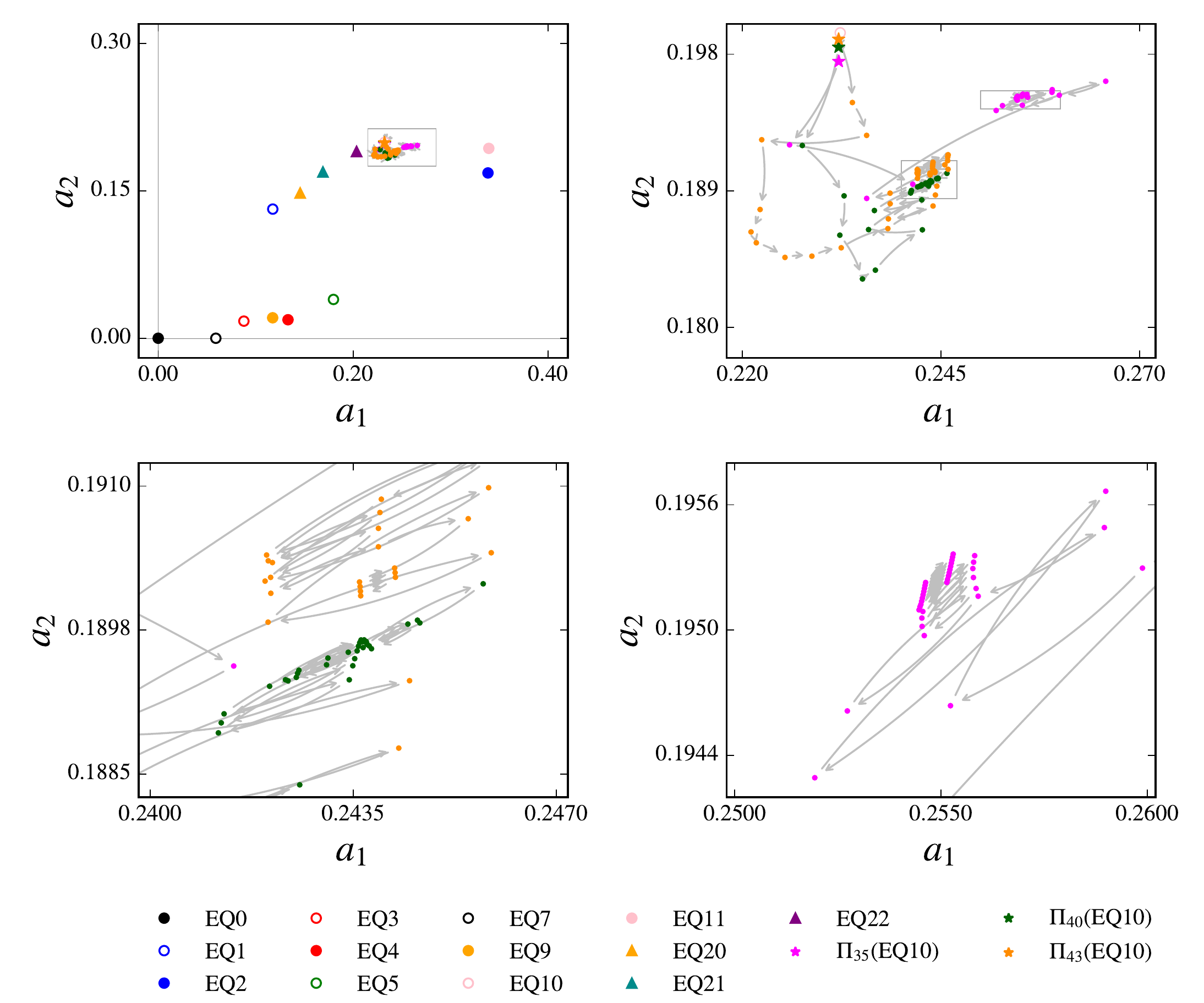}
	\caption{A state space representation of unsuccessful searches initiated from rank projections of EQ10. The plot on the top-right shows the detail of the grey box in the top-left figure and the two plots at the bottom show the detail in the two grey boxes in the top-right plot.}
	\label{fig:search_unsuccessful}
\end{figure}

To demonstrate what unsuccessful searches look like in state space we have plotted the first 40 Newton steps of searches initiated from $\Pi_{38,40,43}($EQ10$)$ in figure \ref{fig:search_unsuccessful}.
The figure shows that the Newton steps oscillate as they near a point, suggesting that the search is stuck near or between manifolds.
Another potential reason for failure to converge may be the prescribed time horizon of the DNS phase when checking \eqnref{eqn:NKH};
in our study a moderate time horizon of $T=20$ was chosen in keeping with \cite{Gibson2009} who chose it to keep the computation time relatively short.
Increasing the time horizon to a much larger value could help in converging the unsuccessful searches and potentially lead to the discovery of other solutions.
Alternatively, the Newton step size could be reduced to prevent overshooting.

% Symmetries
Focussing on the symmetries of the solutions reveals how spatially organised they are.
That is to say, a highly symmetric solution has a very systematic recurrent structure, 
whereas a field which lacks symmetry, such as a turbulent field, lacks exactly repeating structure.
Therefore, highly symmetric solutions are suitable for describing the low-dimensional nature of coherent structures. 
The EQ7-8 family of solutions adhere to the $\Theta$ isotropy subgroup: they are the most symmetric set of equilibria found and as shown in \S\ref{sec:eq_state_space} their unstable manifolds are relatively simple and quickly aim trajectories towards the highly symmetric laminar state.
Flow fields with fewer symmetry constraints can contain more complicated repeating structure.
Equilibria found previously by \cite{Nagata1990}, \cite{Gibson2009}, \cite{Clever1997} and \cite{Waleffe1998, Waleffe2003} in plane Couette flow conform to $\Sigma$ or $\Theta_3$.
Most of the equilibria discovered in our investigation also adhere to those two isotropy subgroups.
However, EQ19 adheres to the subgroup $K$ which exhibits periodicity in the spanwise direction only. 
To our knowledge this is the first time an equilibrium solution in plane Couette flow belonging to this symmetry subgroup has been found.
Interestingly, $K$ is found from a pitchfork bifurcation off the EQ7-8 branch.
As pointed out by \cite{Crawford1991}, when an equilibrium solution branch undergoes a bifurcation, states on the new branch have less symmetry and generally have more complicated dynamics.
In the present work, all bifurcations lead to symmetry-breaking and the loss of symmetry is manifested by the appearance of a new pattern.

Furthermore,
performing the project-then-search method at other Reynolds numbers may find more unknown branches.
Highly symmetric equilibria along the upper-branches of bifurcation curves have the potential to lead to other solution branches because there are symmetries to break.
For example, solutions along the upper-branch of the EQ1-2 bifurcation curve adhere to the isotropy subgroup $\Sigma$ which contains more than one symmetry operation.
Therefore, conducting the project-then-search process at other Reynolds numbers on this branch may lead to the discovery of other symmetry-breaking bifurcations.
The project-then-search method can also be applied to branches of solutions that adhere to the $\Theta_3$ isotropy subgroup of order 1.
Doing so may lead to branches of chaotic solutions that do not obey any symmetries.

Presenting the equilibria in the representation-independent state space in \S\ref{sec:eq_state_space} reveals that the spanwise-shifted equilibria are projected further from the original solution.
Figure \ref{fig:re400_state_space} shows EQ1's unstable manifold and it is clear that a spanwise-shift leads to a larger deviation from its original state: 
EQ1 and $\tau_x$EQ1 are quite close in state space, whereas EQ1 and $\tau_z$EQ1 are quite far apart.
This suggests that the structure within the field varies more in the spanwise direction.
This property is observed in solutions that are far from the laminar state, and EQ1, EQ20, EQ21 and EQ22.
Hence, transitional trajectories that come near the equilibria will be influenced more by the spanwise structure of the equilibria, rather than the streamwise structure.
We argue that the spanwise-shifts play an important role in guiding and shaping transitional trajectories because the state space portraits show that the spanwise-shifts lead to larger deviations in state space.
This is reminiscent of the SSP in which streamwise rolls, which redistribute the mean streamwise-momentum to sustain streaks, experience instability due to `spanwise inflections' that lead to streamwise undulations, ultimately leading back to streamwise rolls via nonlinear self-interaction of the streamwise undulations \citep{Waleffe1997}. 
It is the spanwise inflections that lead to an instability that causes the streamwise-streaks to breakdown to recreate streamwise rolls. 
The breakdown and reformation of coherent structures in the turbulent boundary layer is referred to as bursting in the work of \cite{Viswanath2007}; 
this is an important process that is linked to the onset and maintenance of turbulent flow.
Figure \ref{fig:re270_state_space} shows that EQ18 and EQ27, which belong to the EQ1-2 and EQ20-21-23 solution families respectively, exhibit the same large deviation in the spanwise direction.
Therefore, we suppose that these equilibria may also play an important role in transition and low-Reynolds number dynamics.

The state space portraits also reveal the shape of the laminar attractor.
To our knowledge there are no solutions that have been found to inhabit an unstable manifold that encases EQ1's manifold - this is the most spanwise varying manifold near the laminar solution, closely followed by EQ20.

\section{Conclusions}\label{sec:Conclusions}
% Main result
A systematic exploration of the exact invariant solutions of the NSE was undertaken.
New sets of equilibria were added to the collection of known exact coherent states found by \cite{Nagata1990, Gibson2008, Gibson2009} and \cite{Halcrow2008}.
The equilibria add more structure to the state space of plane Couette flow, and the computation of their unstable manifolds illuminates the pathways that exist in state space.
These pathways can be used to find heteroclinic connections between equilibria and guide turbulent trajectories 
to desired physical states.

% What was the key development in the paper?
The key methodological advance reported here is the computationally cheap method by which initial guesses are generated for the NKH search.
% What is the process?
The resolvent model was used to generate low-rank projections of known equilibria which were then used as seeds for the NKH algorithm with a convergence rate of $96\%$.
If the search succeeded in finding new solutions, then the project-then-search method was applied to the new equilibria, and so on until we obtained a closed set and no new equilibria were discovered.
It is our belief that the project-then-search method's success in finding new solutions is due to the combination of the projections' properties and their location relative to other equilibria in state space.
The energetically dominant physical characteristics of known solutions were maintained in the projections due to the nature of the projection.
% New solutions are near others.
Searches initiated with the projections produced new solutions in the state space neighbourhoods of known solutions since the projections sit near the unstable manifolds of other solutions and the DNS phase of the NKH algorithm allows them to follow the directions of nearby manifolds.
The low computational cost of the projection process makes it effective in obtaining numerous initial states, and thus solutions, from a few previously known ones.
There are countless equilibria to be found; while we restricted ourselves to projections at three previously determined Reynolds numbers we fully expect that the same process at other Reynolds numbers would yield new branches.

% What did not work so well?
In total, $4\%$ of searches initialised by projected flow fields failed to converge to the imposed tolerance despite extending the number of Newton steps. 
The failed searches seemed to be trapped in small regions of state space.
This may be attributed to the small time horizon used for the DNS phase of the searches.
This time horizon can be extended in future studies to avoid failed searches.

% Caveat of method: 
The current approach requires a known solution to find other solutions --- no doubt this a  significant limitation.
% What is your future work?
Nonetheless, many avenues of further work suggest themselves.
The projection method can be modified to truncate the Fourier modes in the homogeneous directions instead of (or as well as) the resolvent modes in the wall-normal direction; 
projections of equilibria at higher Reynolds numbers can be used to initiate searches;
or a different projection basis can be used (for example, resolvent modes defined via a different norm).

% What am I am doing next?
The project-then-search approach can trivially be extended to the other classes of invariant solution.
The case of periodic orbits will be presented in a later paper.\\

% Oscars speech
This work has been supported by the Air Force Office of Scientific Research (European Office of Aerospace Research and Development) under the award FA9550-14-1-0042.
We would like to thank John Gibson for providing his code, equilibrium solutions and helpful comments. 
We would also like to acknowledge Edgar Knobloch for his guidance and Davide Lasagna for his probing questions and useful recommendations.
This work was completed in part at the Kavli Institute for Theoretical Physics, with the support of the National Science Foundation under Grant No. NSF PHY11-25915.

\label{References}
\bibliographystyle{jfm}
\bibliography{references}
\end{document}